\documentclass[fleqn,usenatbib]{mnras}

\usepackage{newtxtext,newtxmath}
\usepackage[T1]{fontenc}

\DeclareRobustCommand{\VAN}[3]{#2}
\let\VANthebibliography\thebibliography
\def\thebibliography{\DeclareRobustCommand{\VAN}[3]{##3}\VANthebibliography}

\usepackage{graphicx}	% Including figure files
\usepackage{amsmath}	% Advanced maths commands
\usepackage{subcaption}
\usepackage{array}
\usepackage{float}
\usepackage{ulem}

\title[Magnetic Fields and High-Mass Star Formation]{The Role of Magnetic Fields in the Formation of High-Mass Star-Forming Cores}

\author[K. S. Klos, I. A. Bonnell, R. J. Smith]{
K. S. Klos,$^{1}$\thanks{E-mail: kk246@st-andrews.ac.uk}
I. A. Bonnell,$^{1}$\thanks{E-mail: iab1@st-andrews.ac.uk}
R. J. Smith,$^{1}$\thanks{E-mail: rjs22@st-andrews.ac.uk}
\\
$^{1}$University of St Andrews, Scotland}

\date{Accepted XXX. Received YYY; in original form ZZZ}

\pubyear{2023}

\begin{document}
\label{firstpage}
\pagerange{\pageref{firstpage}--\pageref{lastpage}}
\maketitle

% HARDCORE
% HydrodynAmical 

\begin{abstract}
Magnetic fields are often invoked as playing a primary role in star formation and in the formation of high-mass stars. We investigate the effect of magnetic fields on the formation of high-mass cores using the 3-dimensional smoothed particle magnetohydrodynamics (SPMHD) code PHANTOM. We follow the collapse of six molecular clouds of mass 1000 M$_{\odot}$, four of which are initially magnetized with mass-to-flux ratios 3, 5, 10 and 100, respectively, and two purely hydrodynamic clouds with varying initial strengths of turbulence. We then apply an in-house clump-finding algorithm to the 3D SPH data in order to quantify the differences in mass and properties of the cores across the degrees of magnetic and turbulent support. We find that although the magnetic fields cause differences in the global cloud evolution, the masses and properties of the cores which form are broadly similar across the different initial conditions. Cores initially form with masses comparable to the initial thermal Jeans mass of the clouds, and then slowly increase in mass with time. We find that regardless of initial magnetization, the fields become dynamically relevant at densities of $\rho > 1\times10^{-17}$ g cm$^{-3}$ - comparable to core densities - and channel material along the field lines, decreasing the stable magnetic Jeans mass, such that the limiting factor for fragmentation is the thermal Jeans mass. We conclude that magnetic fields are not capable of forming and supporting initially high-mass cores against fragmentation. 
\end{abstract}


%%%%%%%%%%%%%%%%% BODY OF PAPER %%%%%%%%%%%%%%%%%%

\section{Introduction}
\label{sec:intro}

High-mass stars ($M > 8 - 10 M_{\odot}$) are a crucial in shaping the interstellar medium (ISM) and driving galactic evolution, as they are responsible for the majority of light, energetic feedback, and chemical evolution \citep{Reynolds1990,Vanbeveren1998,Garay1999,Massey2003,Walch2015,Watkins2019, Mccallum2024}. Despite their importance, their formation mechanism is still hotly debated, as scaling typical low-mass models of star-formation up to high-masses breaks down for a number of reasons \citep{McKee2003, Bonnell2004}.

Magnetic turbulence has often been invoked as a mechanism by which massive cores can form. Numerous observational (e.g., \citet{Crutcher2012, PlanckCollaboration2016, Ching2022}) and numerical works (e.g., \citet{Nakamura2005, Hennebelle2006, Hennebelle2019, Ntormousi2019} have shown that a) magnetic fields are a ubiquitous feature of star-forming regions from cloud to core scales and b) magnetic fields have a significant impact on gas dynamics in these regions, influencing the flow and collapse of gas. As such, the effect of magnetic fields is an important component to consider in the development of a full theory of star-formation.

In the typical description of star-formation, the collapse and fragmentation of gas `clumps' within giant molecular clouds form dense pre-stellar cores. These objects may then accrete further from their surroundings until they reach the densities required to enter the main-sequence \citep{Larson2003, McKee2007}. This process is limited to a few Myr, by which time the gas from the initial cluster is largely dispersed \citep{Lada2003, Smith2011}.

In contrast, high-mass stars obtain central densities high enough to begin hydrogen fusion whilst accretion onto the central protostar is ongoing \citep{Yorke1977,Wolfire1987}. This compounds the issue of high-accretion rates required by high-mass cores in order to complete the star-formation process within the required timescales. Accretion rates up to 100 times higher than typical SF rates are necessary if high-mass stars form from the same initial core masses as low-mass stars \citep{Zinnecker2007}.

Furthermore, high-mass stars are preferentially found within the dense central regions of young stellar clusters, displaying strong mass-segregation \citep{Gouliermis2004,Bontemps2010}. Mass-segregation is a natural outcome of dynamical interactions between stars within the cluster, slowing down the most massive objects and causing them to sink to the center of the cluster potential \citep{Binney1987}. This process would explain the apparent lack of mass-segregation in some massive star-forming regions \citep{Gennaro2017}; there has simply not been enough time for the stellar components to relax into a mass-segregated configuration. However, observations have shown numerous mass-segregated clusters with ages significantly lower than predicted timescales for dynamical segregation \citep{Bonnell1998,Hillenbrand1998,Gouliermis2004}. This implies that the distribution of masses within clusters is likely a product of their formation mechanism. Understanding the origin of mass-segregation in clusters is likely a key part to developing a theory for massive star-formation.

These differences have led to the emergence of two main descriptions of high-mass star-formation. In the `competitive accretion' scenario \citep{Bonnell2001}, a collapsing cloud initially fragments to form multiple low-mass cores. Those which find themselves at the centers of local potential wells can accrete more efficiently from the gas reservoir than less fortunate cluster members on the outskirts. These `lucky' cores are then able to reach higher masses within comparable core-formation timescales. This description naturally provides an explanation for young mass-segregated clusters, as the highest mass objects necessarily form in the central regions of their cluster. However, a small yet non-negligible number of `lone' massive stars have been observed \citep{Bestenlehner2011, Selier2011, Bressert2012, Law2022}. These stars do not appear to have been dynamically ejected from their original clusters, raising the question `did these stars form in isolation'?

In contrast, the `\textit{Turbulent Core}' model, proposed by \citet{McKee2002,McKee2003}, suggests that the fragmentation of collapsing gas clouds is counter-balanced by turbulent pressure support. Cores form with much higher masses and densities, which, once they become unstable, collapse on shorter free-fall timescales \citep{McKee2002}. This in turn leads to higher accretion rates onto the central protostar, as required by formation timescale constraints. Such a scenario would be able to explain the formation of isolated massive stars. Unfortunately, subsequent numerical and observational studies have broadly confirmed that hydrodynamical turbulence in clouds is not of the magnitude required to prevent fragmentation \citep{Dobbs2005, Bontemps2010, Pillai2011, Tackenberg2012, Beuther2013, Beuther2015, Bihr2015, Pillai2019, Beuther2021}.

However, as we have already discussed, magnetic fields may also support material in the ISM from collapse. The degree of magnetic support in a cloud can be quantified by the mass-to-flux ratio. This is just the total mass $M$ divided by the magnetic flux threading the region, $\Phi$. It is often quoted in terms of the critical mass-to-flux ratio, $(M/\Phi)_{\mathrm{crit}}$ given by \citet{Mestel1999, MacLow2004}
\begin{equation}
    \frac{M}{\Phi}_{\mathrm{crit}} = \frac{2c_{1}}{3} \sqrt{\frac{5}{G\pi \mu_{0}}}
    \label{MFcrit}
\end{equation}
where $c_{1} = 0.53$, $\mu_{0}$ is the permeability of free space, and $G$ is the gravitational constant \citep{Mouschovias1976}. On cloud-scales, magnetic support is often comparable to the self-gravity of the cloud, giving subcritical (i.e., $M/\Phi < 1$) ratios, preventing collapse. However, on smaller core-scales mass-to-flux ratios show a transition to supercritical ($M/\Phi > 1$) values facilitating collapse. Typical observed values span $\sim 3 - 5$ \citep{Ching2022}.

The support from the field is still significant in this regime, however. Observationally, `hourglass' core morphologies imply that regions perpendicular to the direction of the field are collapsing at a lower rate than material flowing parallel to the field \citep{Kirby2009, Dotson2010, Beltran2019}. Numerical simulations of magnetized collapsing cores demonstrate a tendency to form higher-mass, and fewer, cores \citep{Commercon2011,Hennebelle2011,Rosen2020}. The star-formation rate is reduced, and field provides additional support against fragmentation \citep{Hennebelle2011}. 

However, the majority of numerical studies in recent years have focused on the evolution of magnetized cores once they have already formed, typically on size and mass scales $\sim 0.1$ pc and $\sim 10 - 100 M_{\odot}$ respectively (e.g., \citet{Rosen2020, Rosen2022}). The purpose of the work herein is to investigate if the inclusion of magnetic support facilitates the formation of high-mass cores from the collapse of a larger, more extended clump in the first place. We simulate the collapse of a six 1000 $M_{\odot}$, $R = 1$ pc `clumps', for which the simulations and initial conditions are described in section \ref{sec:method}. We then analyze the properties of `cores' using a clump-finding algorithm described in section \ref{subsec:CF} based on the predictions of the \textit{Turbulent Core} model. In section \ref{sec:results} we summarize the key results from these simulations and the clump-finding analysis. We discuss these results in the context of magnetic fields and star-formation within section \ref{sec:discussion}, and conclude in section \ref{sec:conclusions}

\section{Simulations}
\label{sec:method}
\subsection{SPMHD Code}
\label{subsec:SPMHD}
In this work we make use of the three-dimensional Smoothed Particle Hydrodynamics (SPH) code PHANTOM, including self-gravity, ideal magnetohydrodynamics (MHD), accreting sink particles \citep{Price2018phantom}. PHANTOM solves the discretized forms of ideal MHD, as defined in \citep{Price2012}, given by
\begin{equation}
\begin{aligned}
    \rho_{a} = \sum_{b} m_{b}W_{ab}(h_{a}); && h = h(\rho)
    \label{eq:rho}
\end{aligned}
\end{equation}
\begin{equation}
\begin{aligned}
    \frac{\mathrm{d}v^{i}_{a}}{\mathrm{d}t} = & \sum_{b} m_{b} \left[ \frac{S^{ij}_{a}}{\Omega_{a}\rho_{a}} \nabla_{a}^{i}W_{ab}(h_{a}) + \frac{S^{ij}_{b}}{\Omega_{b}\rho_{b}} \nabla_{a}^{i}W_{ab}(h_{b}) \right] \\
    & + \Pi_{\mathrm{shock}}^{a} + f_{\mathrm{divB},a}^{i}  +a_{\mathrm{sink-gas}}^{i} - \nabla^{i} \Phi_{a}
    \label{eq:mot}
\end{aligned}
\end{equation}
\begin{equation}
\begin{aligned}
    \frac{\mathrm{d}}{\mathrm{d}t} \left(\frac{B_{a}^{i}}{\rho_{a}}\right) = & - \sum_{b} m_{b}(v_{a}^{i} - v_{b}^{i})\frac{B_{a}^{i}}{\Omega_{a}\rho_{a}^{2}} \cdot \nabla^{i} W_{ab}(h_{a}) \\
    & - \sum_{b} m_{b} \left[\frac{\psi_{a}}{\Omega_{a}\rho_{a}^{2}} \nabla_{a}W_{ab}(h_{a}) + \frac{\psi_{b}}{\Omega_{b}\rho_{b}^{2}} \nabla_{a}W_{ab}(h_{b})\right] \\
    & + \frac{1}{\rho_{a}}\mathbf{\mathcal{D}}_{\mathrm{diss}}^{a}
    \label{eq:ind}
\end{aligned}
\end{equation}
\begin{equation}
\begin{aligned}
    \nabla^{2}\Phi_{a} = 4\pi G\rho_{a}
    \label{eq:pot}
\end{aligned}
\end{equation}
where the MHD stress tensor, $S^{ij}$, is defined by
\begin{equation}
\begin{aligned}
    S^{ij} \equiv \left(P + \frac{1}{2\mu_{0}}B^{2}\right)\delta^{ij} + \frac{1}{\mu_{0}}B^{i}B^{j}
    \label{eq:stress}
\end{aligned}
\end{equation}
where $\rho$ is the density, $m$ is mass, $W_{ab}$ is the smoothing kernel, which is dependent on the smoothing length, $h_{a}$. $v$ is velocity, $\frac{\mathrm{d}}{\mathrm{d}t} \equiv \frac{\partial}{\partial t} + \mathbf{v} \cdot \nabla$ is the Lagrangian derivative, $\Omega_{a}$ accounts for the gradient of the smoothing length and is defined as
\begin{equation}
\begin{aligned}
    \Omega_{a} \equiv \left[1 - \frac{\partial h_{a}}{\partial \rho_{a}} \sum_{b} m_{b}\frac{\partial W_{ab}(h_{a})}{\partial h_{a}}\right] 
\end{aligned}
\end{equation}
$\Pi_{\mathrm{shock}}^{a}$ is a shock energy dissipation term, $f_{\mathrm{divB}, a}^{i}$ is a source term introduced to prevent the tensile instability which occurs in Smoothed Particle MHD (SPMHD), $a_{\mathrm{sink-gas}}^{i}$ includes the effect of sink particles on gas motions, $P$ is pressure, $\mu_{0}$ is the permeability, $B$ is the magnetic field, $\delta^{ij}$ is the Kronecker delta, $\psi$ is a scalar value used into control the error in the divergence of the magnetic field, and $\mathcal{D}_{\mathrm{diss}}$ is magnetic dissipation, $\Phi$ is the gravitational potential. For a more in-depth discussion of these terms see \citet{Price2018phantom}.

\subsubsection{Equation of State}
\label{subsubsec:EOS}
A barotropic equation of state is used to model the evolution of the cloud \citep{Larson1969}. It is described by
\begin{equation}
\begin{aligned}
    P=\begin{cases}
        c^{2}_{\mathrm{s}} \rho, & \rho < \rho_{\mathrm{a}} \\
        c^{2}_{\mathrm{s}}\left(\frac{\rho}{\rho_{\mathrm{a}}}\right)^{\gamma_{3}}, & \rho_{\mathrm{a}} < \rho < \rho_{\mathrm{c}} \\
        c^{2}_{\mathrm{s}}\left(\frac{\rho_{c}}{\rho_{\mathrm{a}}}\right)^{\gamma_{3}}\left(\frac{\rho}{\rho_{\mathrm{c}}}\right)^{\gamma_{1}}, & \rho_{\mathrm{c}} < \rho < \rho_{\mathrm{d}} \\
        c^{2}_{\mathrm{s}}\left(\frac{\rho_{d}}{\rho_{\mathrm{c}}}\right)^{\gamma_{1}}\left(\frac{\rho_{c}}{\rho_{\mathrm{a}}}\right)^{\gamma_{3}}\left(\frac{\rho}{\rho_{\mathrm{d}}}\right)^{\gamma_{2}}, & \rho > \rho_{\mathrm{d}}
    \end{cases}
\end{aligned}
\end{equation}
where the threshold densities correspond to $\rho_{\mathrm{a}} = 10^{-16}$ g cm$^{-3}$, $\rho_{\mathrm{c}} = 10^{-14}$ g cm$^{-3}$, $\rho_{\mathrm{d}} = 10^{-10}$ g cm$^{-3}$, and the adiabatic indices correspond to $\gamma_{1} = 7/5$, $\gamma_{2} = 11/10$, and $\gamma_{3} = 5/3$. $\rho_{\mathrm{c}}$ and $\rho_{\mathrm{d}}$ are the same as used by \citet{Lewis2015} for modeling the collapse and formation of prestellar cores. We have introduced the second condition; $\rho_{\mathrm{a}} < \rho < \rho_{\mathrm{c}}$, in order to slow the formation of sink particles in the simulation (see section \ref{subsubsec:sink}). We do this for several reasons; firstly the insertion of a sink particle effectively removes the magnetic field from the region enclosed by it's accretion radius \citep{Wurster2016}, reducing the impact of the cloud magnetization as the simulation evolves, and secondly to minimize the impact of sink-gas interactions on the interpretation of core morphologies at later times. The threshold density $\rho_{\mathrm{c}} = 10^{-16}$ g cm$^{-3}$ is chosen as it is a factor of 10 below the sink creation threshold, allowing high-density gas to build up before immediately attempting sink creation.

\subsubsection{Sink Particles}
\label{subsubsec:sink}
Sink particles \citep{Bate1995} are used to model the evolution of the cloud beyond the point at which the gas begins to fragment. We choose $\rho_{\mathrm{crit}} = 10^{-15}$ g cm$^{-3}$ as the sink creation threshold. As discussed above, whilst the inclusion of sink particles has benefits with regards to simulation efficiency, the impact of their interactions on the gas structures in the cloud prove problematic in this study, especially with regards to the discussion of morphologies at later times. For this reason the sink particle parameters we have chosen aim to balance between simulation efficiency and a more realistic treatment of the gas. The sinks are given accretion radii $h_{\mathrm{acc}} = 0.01$ pc, and merge conditionally within $3 h_{\mathrm{acc}}$ of each other and unconditionally within $2 h_{\mathrm{acc}}$, the latter of which is the default minimum recommendation when using PHANTOM. In order for sinks to be inserted into the simulation a number of physical and numerical checks must be passed (see \citet{Price2018phantom} for detailed explanation of how sinks are treated within PHANTOM).

\subsection{Initial Conditions}
\label{subsec:initcond}
We performed a set of six simulations following the collapse of a spherical 1000$M_{\odot}$ cloud. The sphere contained 8 $\times$ 10$^{7}$ particles giving a particle mass resolution of $m_{\mathrm{p}} = 1.25 \times 10^{-5}$ M$_{\odot}$, corresponding to a minimum resolvable fragmentation mass of $2M_{\mathrm{kern}} = 1.45 \times 10^{-3}$ M$_{\odot}$, where $M_{\mathrm{kern}}$ is the kernel mass. The clouds where modelled by initially uniform density spheres with a radii $R_{\mathrm{cl}} = 1$ pc, giving an initial density of $1.61 \times 10^{-20}$ g cm$^{-3}$, and a free-fall time of $t_{\mathrm{ff}} = \sqrt{3\pi/32G\rho} = $ 0.52 Myr. The temperature of the clouds was set to 10 K, with a sound speed $c_{\mathrm{s}} = 2.189 \times 10^{4}$ cm s$^{-1}$. They also all have an initial rotation along the $z$-axis of $\Omega = 10^{-14}$ rad s$^{-1}$. There are four initially magnetized clouds with mass-to-flux ratios (in units of the critical value) $M/\Phi =$ 3, 5, 10, and 100, corresponding to magnetic field strengths of $B_{\mathrm{c}} =$ 45.5, 27.3, 13.7, and 1.37 $\mu$G respectively. The fields were uniform and initially aligned with the rotation axis, and we assume that on these size scales, we do not have to take into account non-ideal MHD effects \citep{Wurster2016}. We also compare to two non-magnetized clouds.

The clouds are seeded with an initially turbulent field which is allowed to decay over the course of the simulation. Turbulence is calculated using a divergence-free random Gaussian velocity field, such that $P(k) \propto k^{-4}$ \citep{Ostriker2001, Bate2003}. Particle velocities are interpolated from the grid such that $\langle v_{\mathrm{turb}} \rangle = \mathcal{M}c_{\mathrm{s}}$, where $\mathcal{M}$ is the mach number. Five of the six clouds start in virial equilibrium such that the sum of the kinetic energy, $T$, and potential energy $U$ gives $2T + U = 0$. We also have cloud HDC for which the initial conditions are subvirial, such that it collapses virtually unimpeded. These conditions give turbulent mach numbers of $\mathcal{M} =$ 6.59 and 4.66 respectively. Initial field strengths, energy ratios, as well as names for each cloud are summarized in table \ref{tab:names}. The cold cloud was included to provide a relative comparison of the influence of the initial turbulence on the evolution of the cloud along side the magnetic field.

The initial clouds are contained within a uniform low-density box, with a density contrast of 30:1 with the initial cloud, following previous studies of star-formation in a magnetized medium (e.g., \citet{Wurster2016,Lewis2017,Wurster2021}. The box has quasi-periodic boundaries, allowing the SPH particles to interact hydrodynamically, but not gravitationally, across the box. The length of the box sides are $L = 4R_{\mathrm{cl}}$

\begin{center}
\begin{table}
\centering
\begin{tabular}{ | m{5em} | m{3cm}| m{1cm}|} 
  \hline
  \textbf{Cloud ID} & $M/\Phi$ ($(M/\Phi)_{\mathrm{crit}})$ & $T/U$ \\
  \hline
  MHD3 & 3 & 0.5 \\ 
  MHD5 & 5 & 0.5 \\ 
  MHD10 & 10 & 0.5 \\ 
  MHD100 & 100 & 0.5 \\ 
  HDV & N/A & 0.5 \\ 
  HDC & N/A & 0.4 \\ 

  \hline
\end{tabular}
\caption{Summary of cloud names and their initial magnetization (given by the mass-to-flux ratio), and their virial state. All clouds are initially virialized, except HDC which is initially sub-virial.}
\label{tab:names}
\end{table}
\end{center}

\subsection{Clump-finding}
\label{subsec:CF}
In order to characterize the masses of cores within our simulations, we have developed a clump-finding algorithm based on the work by \citet{Wurster2023}, which sorts particles by density within the simulation, and compares their local potential and kinetic energies to identify bound structures. However, as discussed in \citep{Smith2009}, using density as a criterion for identifying 3D simulated cores can be problematic, as it does not necessarily trace the location of local potential minima.

\begin{figure}
    \centering
    \includegraphics[width=\columnwidth]{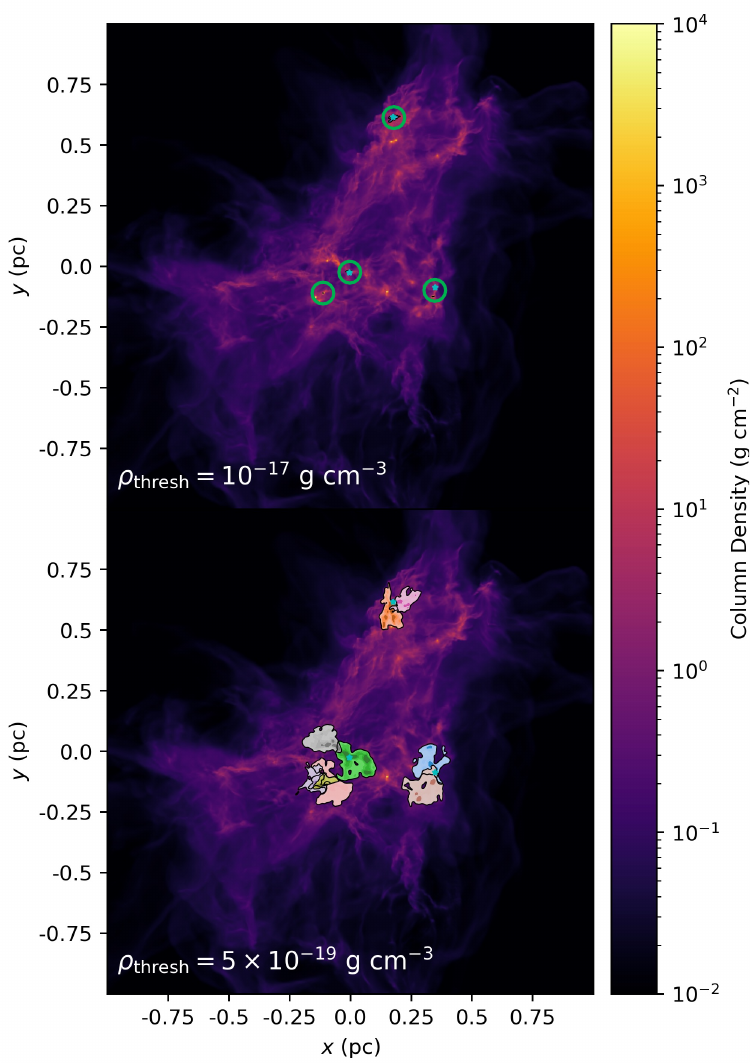}
    \caption{Column density plot showing the location of cores found in an example cloud by the clump-finding algorithm using two different density thresholds. The top panel shows cores found for a higher density threshold of $\rho_{\mathrm{thresh}} = 1 \times 10^{-17}$ g cm$^{-3}$, and the bottom panel shows cores found with a lower threshold of $\rho_{\mathrm{thresh}} = 5 \times 10^{-19}$ g cm$^{-3}$. Green circles indicate the location of cores in the top panel as they are not easily visible. Sinks are shown as cyan stars.}
    \label{fig:corelocscomp}
\end{figure}

Instead, we use a combined approach, using both density and potential to identify the location of cores. First, we locate particles located in density structures, setting a density threshold, $\rho_{\mathrm{thresh}}$, to be considered as part of a core. The choice for this value can have significant effects on the masses and morphologies of cores identified by the clump-finding algorithm. Higher density cuts lead to lower core masses, and may even split cores into two distinct objects, as shown in figure \ref{fig:corelocscomp}. For the top panel we see that the density threshold is very high, leading to very small cores, which are virtually invisible in the figure. For the lower density threshold, however, we get much more extended structures, and a greater number of cores. We chose a physically motivated density threshold for use in the clump-finding algorithm. Observations typically infer core sizes on the scale of 0.01 - 0.1 pc, with masses up to 10$^{2}$ M$_{\odot}$ (e.g., \citet{Morii2023}), giving densities of order $10^{-17}$ g cm$^{-3}$ assuming uniform density. We relax this slightly, and lower the density threshold to $5 \times 10^{-18}$ g cm$^{-3}$, in between the two thresholds given in figure \ref{fig:corelocscomp}, as cores show centrally condensed density profiles (e.g., \citet{EvansII1999, WardThompson1999}, with material on the outer edges likely to be lower than this approximate value. However, as the purpose of this study is to compare the variation in trends, and not exact values, between cores in the different clouds, the conclusions drawn from the results should not be significantly altered by the choice of density threshold.

Full details of the clump-finding algorithm can be found in the appendix, but the basic requirements for core identification are as follows. Once the density threshold is applied to the cloud material, local potential minima are identified within the cloud. Particles identified at the peaks of these minima are assigned as initial core `seeds', from which the algorithm ``builds'' cores based on the requirements that these structures are virialized; i.e., $2T + U \leq 0$, they form one connected structure, that they are locally converging, and that they are bound when considering the contributions of thermal and magnetic energy, $T + U + E_{\mathrm{thr}} + E_{\mathrm{mag}} \leq 0$. We also require that each core is large enough to resolve fragmentation, i.e., the number of particles in a core is greater than two times the kernel mass.

\section{Results}
\label{sec:results}
\subsection{Cloud Evolution}
\label{subsec:cloudev}
\begin{figure*}
    \centering
    \includegraphics[width=\linewidth]{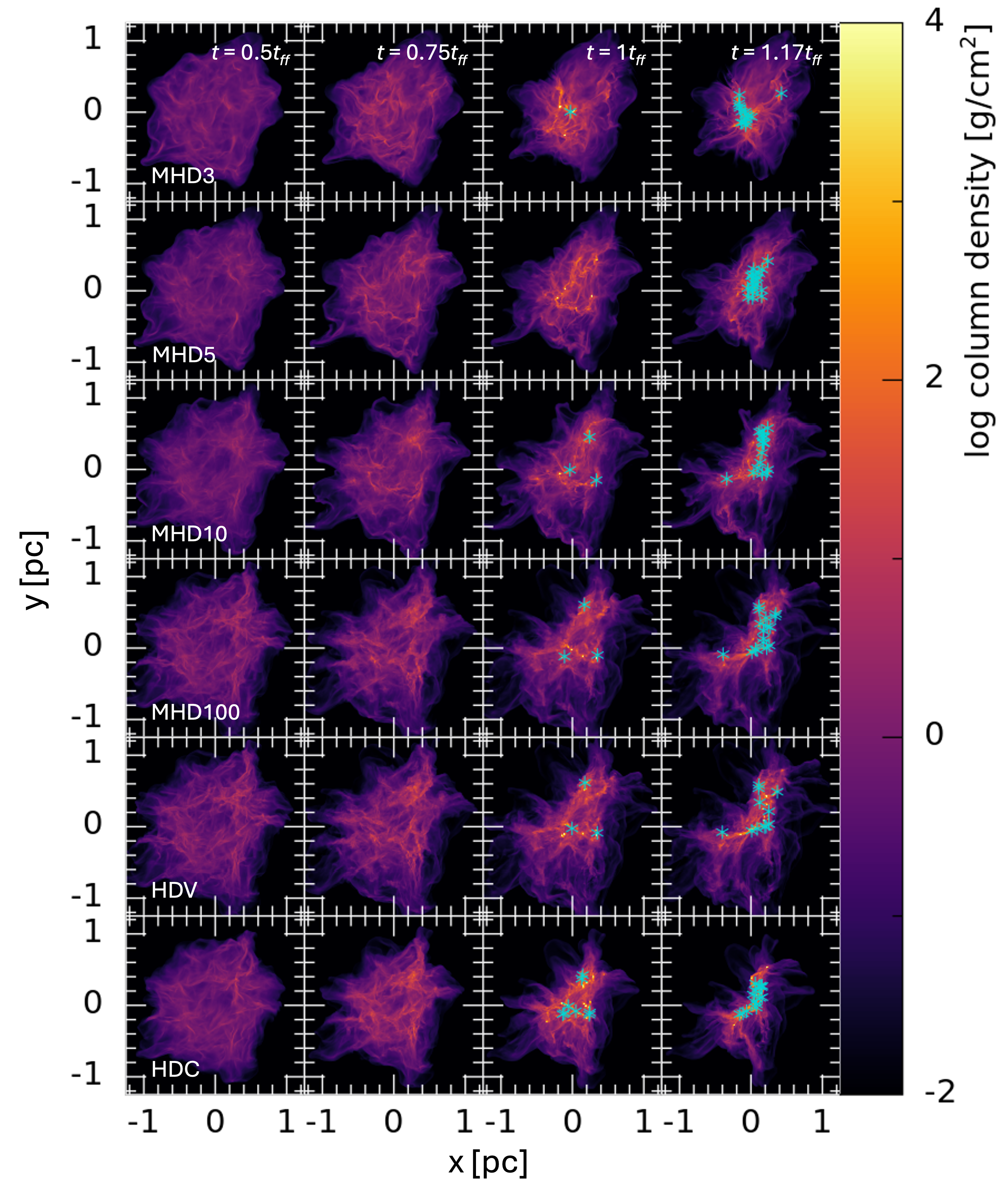}
    \caption{Column density plots showing the evolution of the cloud at different times up until one free-fall time. Visually, the highly-magnetized clouds demonstrate more centrally condensed density profiles, with less substructure formation compared to clouds with weaker/no fields. Furthermore, as the strength of the field decreases, the number of sink particles (cyan stars) present for a given time increases. This further indicates the delay in density evolution caused by the presence of magnetic fields.}
    \label{fig:columndense3vol}
\end{figure*}

As discussed in section \ref{subsec:initcond}, the clouds all begin as uniform density spheres, which collapse over the course of 1.17 $t_{\mathrm{ff}}$. We quantify time in terms of the gravitational free-fall time, which is the same for all of the clouds as it only depends on the initial density (see section \ref{subsec:initcond}). However, as will be discussed below, the presence of magnetic fields alters the actual collapse timescale of the clouds, making $t_{\mathrm{ff}}$ less suitable for comparing cloud evolution. We simulate the clouds slightly past one free-fall time, as it allows for structures to form and evolve past the typical cloud collapse timescale, whilst also stopping early enough that the formation and subsequent evolution of sink particles have not drastically altered the appearance of the cloud. Figure \ref{fig:columndense3vol} visualizes the cloud at three points in time, $0.5 t_{\mathrm{ff}}$, $0.75 t_{\mathrm{ff}}$, and $1 t_{\mathrm{ff}}$, for the six initial conditions.
At $0.5 t_{\mathrm{ff}}$, the six clouds appear broadly similar. They retain the outline of their initial spherical shape, and have internal structure driven by the initial turbulence. We see that with the exception of MHD3, the locations of these structures appear to be predominantly in the upper right, and lower left portions of the cloud, forming a loose ring around less dense central region. In contrast, MHD3 at $0.5 t_{\mathrm{ff}}$ shows a more uniform distribution of over-densities, with no clear preferred location.

As the clouds evolve, the differences in these structures becomes more apparent. Clouds with low/no initial magnetization maintain a less dense central region, with the upper right sections continuing to condense into what visually may be identified as cores, and this is confirmed by the formation of sink particles in those regions by $1 t_{\mathrm{ff}}$. Clouds MHD10, MHD100, and HDV show the most similar evolutions, each forming three sink particles (green points in figure \ref{fig:columndense3vol}) by $1 t_{\mathrm{ff}}$ in virtually the same locations. This implies that on the larger cloud scale, the impact of the magnetic field is minimal, especially in the case of MHD100. Generally, the dense ring identified at earlier times continues to condense, and there are substructures within these filaments that appear to be forming cores. Overall, this leads to a more irregular cloud structure, with potential cores spread out from each other.

\begin{figure}
    \centering
    \includegraphics[width=\columnwidth]{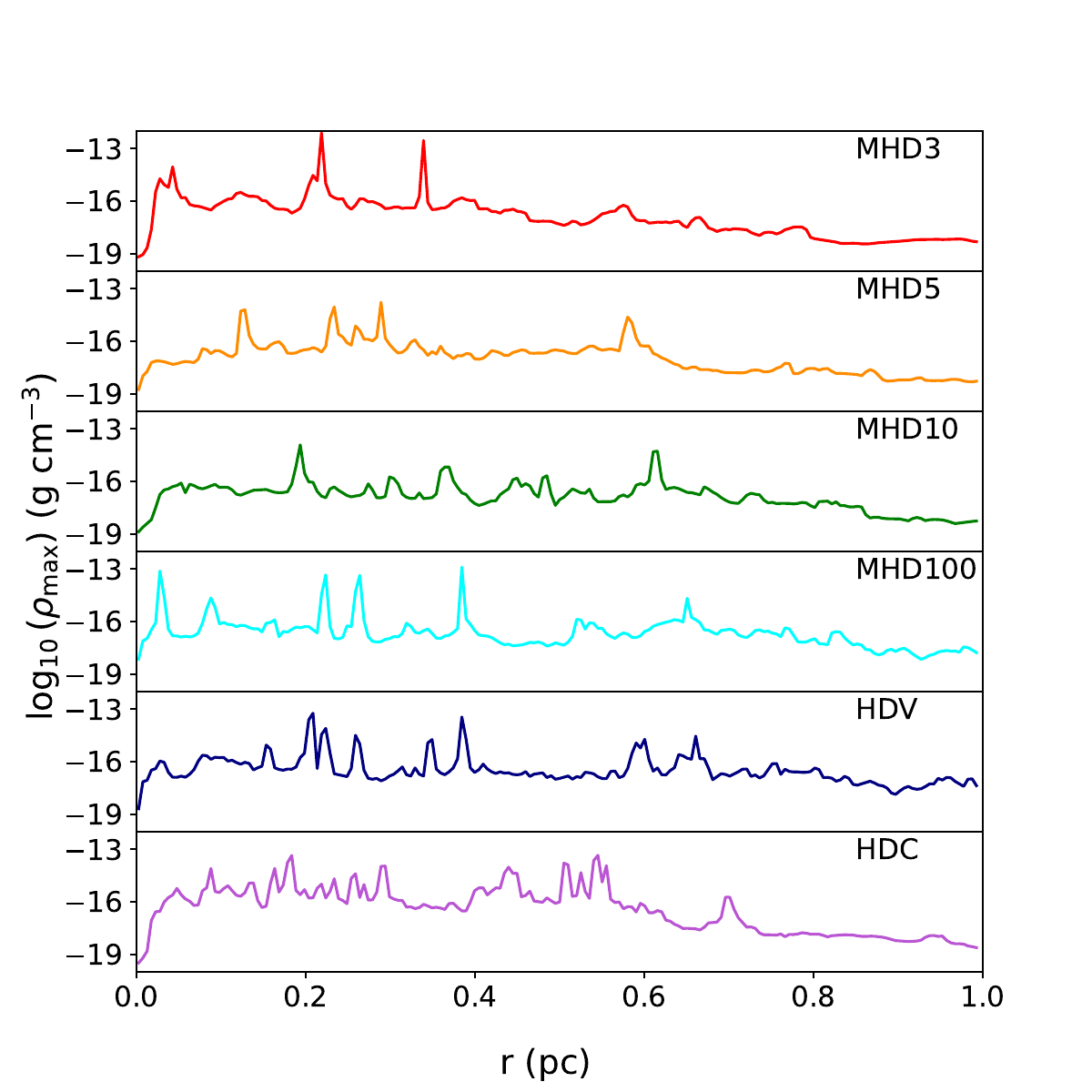}
    \caption{Peak density plots showing the maximum density per radius bin throughout the cloud, at $t = 1 t_{\mathrm{ff}}$. For MHD3, the cloud has two clear peaks at $r \sim 0.2$ and $r \sim 0.4$, as well as a central `bump' which is also forming a peak. As initial magnetization decreases, the location of peaks becomes much more spread out throughout the cloud radius, and the number of peaks also increases, from about three for MHD3 to closer to ten in the cases of HDV and HDC.}
    \label{fig:maxdbins}
\end{figure}

\begin{figure}
    \centering
    \includegraphics[width=\columnwidth]{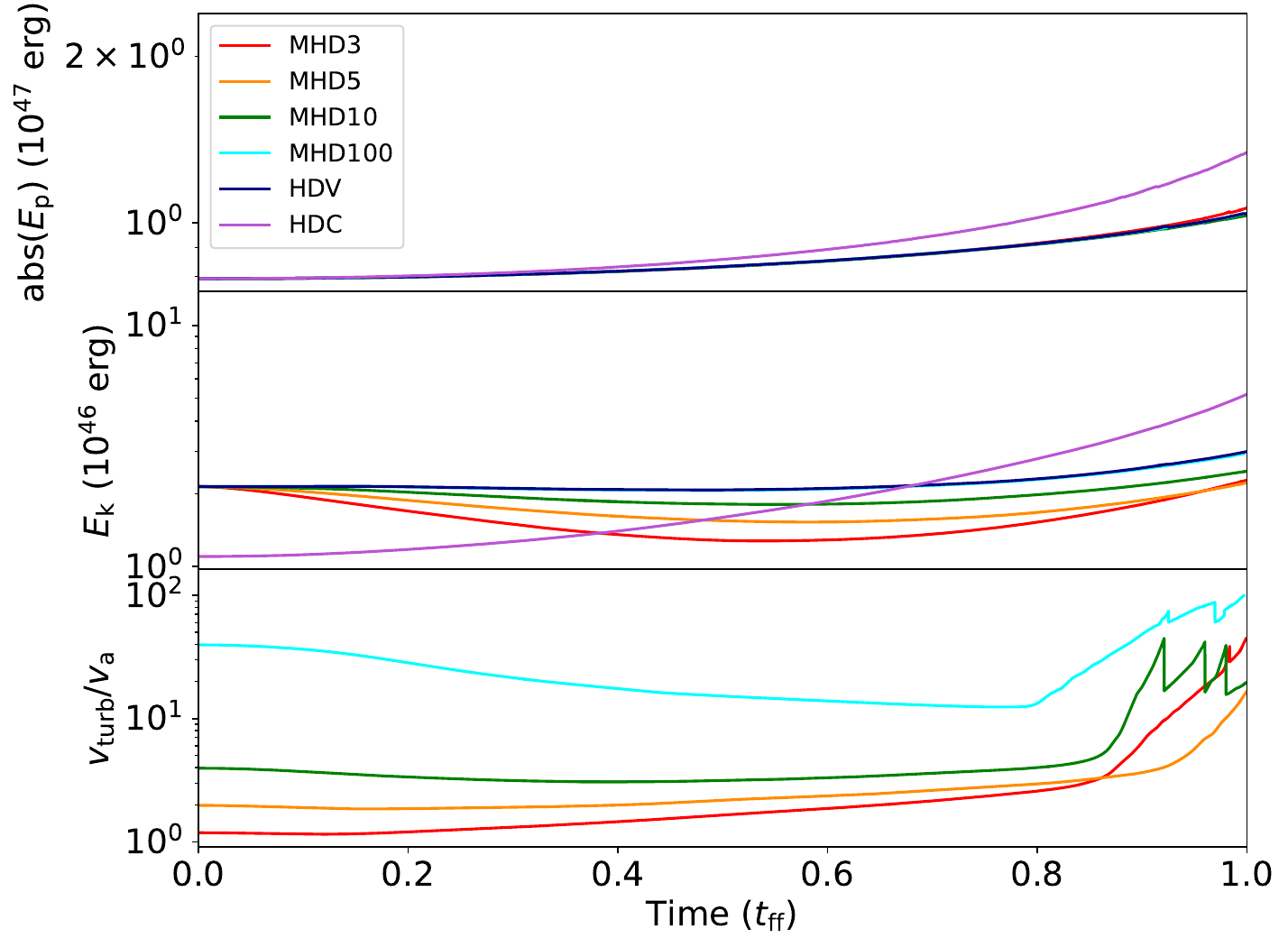}
    \caption{Evolution of potential (top panel), and kinetic (second panel) energies, including boundary particles. The bottom panel shows the evolution of the mean Alfven mach number for the magnetized clouds. For virialized clouds, we see the evolution of potential energy proceed similarly up until approximately one free-fall time. In contrast, the cold cloud (HDC) contracts much more rapidly, owing to the lack of turbulent support. Low/non-magnetized clouds show a gradual increase in the global kinetic energy, whereas the highly magnetized clouds first show a dip, followed by an increase. In the bottom panel, we see that for MHD3, the contribution from magnetic turbulence is comparable to that of the initial turbulence supplied to the cloud, explaining the relative lack of turbulent structure seen in this cloud compared with the less magnetized clouds. The spikes in the lower panel for MHD10 and MHD100 are caused due to the sudden change in $\rho_{\mathrm{ave}}$ for the cloud, as sink particles are created.}
    \label{fig:alf_ekpot}
\end{figure}

\begin{figure}
    \centering
    \includegraphics[width=\columnwidth]{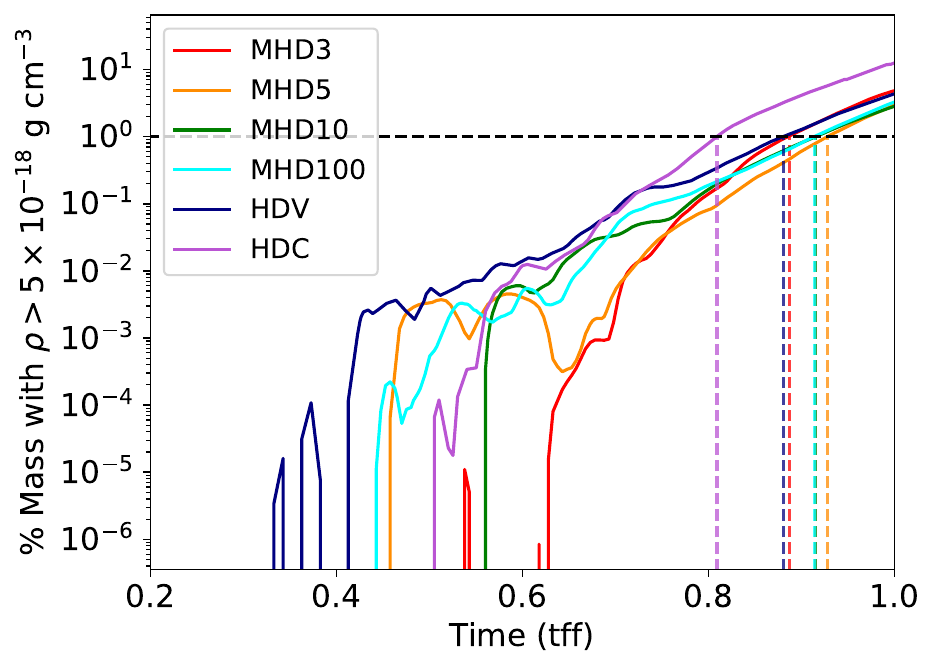}
    \caption{Evolution of mass above the core density threshold as a function of time for each cloud. The black dashed line indicates a mass-fraction of 1\% (equivalent to $\sim$ 10 M$_{\odot}$) above the core threshold density; $\rho_{\mathrm{thresh}} = 5 \times 10^{-18}$ g cm$^{-3}$. For the same point in physical time, we see clouds are at significantly different points in their physical evolution, with cloud HDC (non-magnetized, cold) reaching a fraction of 1\% at a time of $t \approx 0.81 t_{\mathrm{ff}}$. Comparatively, cloud MHD5 only reaches this point at time $t \approx 0.92 t_{\mathrm{ff}}$.}
    \label{fig:massfracevol}
\end{figure}

However, in the case of MHD3, as well as MHD5 to a lesser extent, the collapse proceeds much more centrally. In this case, mass accumulates in the central regions, and the substructure is more homogeneous. Unlike the low-magnetization clouds at $1 t_{\mathrm{ff}}$, the MHD3 cloud forms a sink particle almost exactly at the cloud center. Figure \ref{fig:maxdbins} also illustrates this point. We see that plotting the highly magnetized clouds, the locations of density spikes lie much more centrally than for clouds of lower/no magnetization. The density spikes of MHD100 trace those of HDV, indicating the negligible effect of the field on the cloud collapse. For MHD10, the shape and number of density spikes is closer to that of MHD5, but shifted slightly outwards. In general, the field acts to support the cloud against gravitational collapse, suppressing the formation of substructure. It is only when material becomes self-gravitating at the center of the cloud that collapse can occur.

The similarity in structures is a result of the turbulent velocity field imprinted on the cloud. We can see this more closely when looking at the magnetized clouds. Figure \ref{fig:alf_ekpot} shows the evolution of global cloud potential and kinetic energies, as well as the mean Alfven mach number in the magnetized clouds. In the top panel we see that for all the virialized clouds, the global evolution of potential energy is virtually the same for $t < 0.9 t_{\mathrm{ff}}$. This indicates that the magnetic field is having little effect on the global collapse of the cloud. However, we see significant differences in the kinetic energy evolution for the magnetized clouds. This can be understood from the Alfven speed. For MHD3, MHD5, and MHD10, the ratio of the turbulent to Alfven velocities are approximately 1, 2, and 4. As the turbulence is already supersonic, with $\mathcal{M} =$ 6.59, these three clouds have trans- and supersonic Alfven waves, which can drive magnetic shocks into the cloud. These shocks dissipate the kinetic energy of the cloud, and act to erase the initial turbulent structure. This is shown most clearly for MHD3 where $v_{\mathrm{turb}}/v_{\mathrm{a}} \sim 1$. It is only at later times when the Alfven speed is lower that the kinetic energy starts to rise again. For MHD100, the mean Alfven velocity remains below the sound speed throughout the simulation. The formation of structure is totally dominated by the initial turbulent field, and the cloud evolves in an almost identical manner to HDV.

The differences in cloud evolution mean that by the time one free-fall time is achieved, not only are the shapes and locations of structure different, as well as the amount of material available at different densities. Figure \ref{fig:massfracevol} shows the amount of material in the cloud above a density threshold, $\rho_{\mathrm{thresh}}$, as a function of free-fall time. We clearly see that low/non magnetized clouds initially form more high-density material than clouds with higher magnetizations. As the simulations progress, we see the magnetized clouds condense at similar rates, with the slight exception of MHD3. When comparing to figure \ref{fig:alf_ekpot}, MHD3 shows an increase in the formation of material at densities above $\rho_{\mathrm{thresh}}$ in line with a sharp drop in the Alfven speed. We can see the effect of these processes in the results from the clump-finding algorithm, especially in the masses (section\ref{subsec:masses}) and numbers (section \ref{subsec:CMF}).

\subsection{Core Masses}
\label{subsec:masses}
\begin{figure}
    \centering
    \includegraphics[width=\columnwidth]{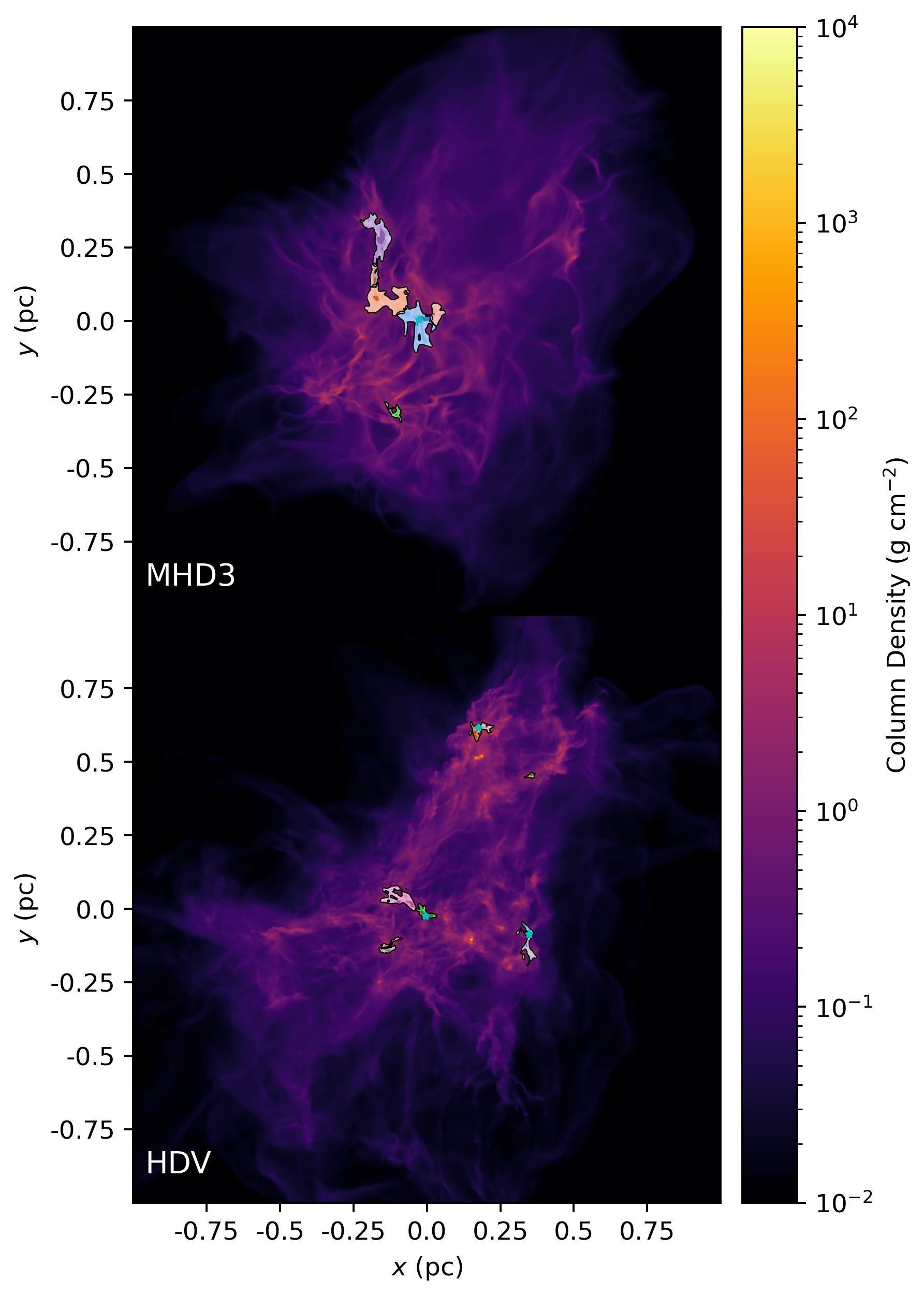}
    \caption{Column density plot showing the location of cores for cloud MHD3 (top), and cloud HDV (bottom). The cores for MHD3 are much more centrally located, and extended than the cores found for HDV. However, when comparing the masses and densities with figure \ref{fig:core_mass1tff}, we see that the masses in each cloud are comparable, whilst the core densities are generally higher for HDV.}
    \label{fig:corelocscompMHD3HDB}
\end{figure}

\begin{figure*}
    \centering
    \includegraphics[width=\linewidth]{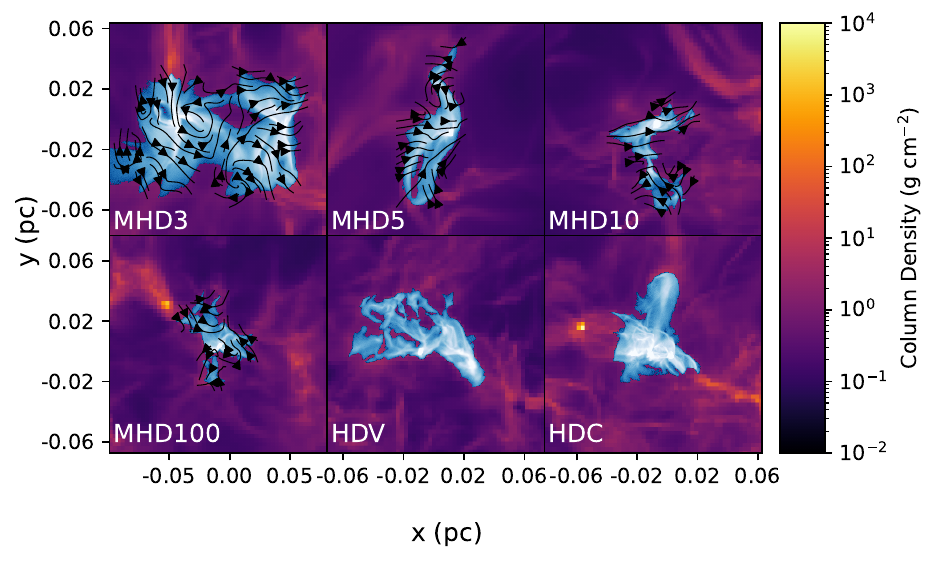}
    \caption{Zoom-ins the most massive gas cores for each cloud. Each panel is centered on the core center-of-mass. For the magnetized clouds we also include the streamlines of the magnetic field in the cores, which show the field has been warped by the collapse. None of the cores have a classical spherical morphology, owing to the interactions between the field, turbulence, and collapse. The core for MHD3 has a mass of 17 M$_{\odot}$ is the most extended core, and shows the smoothest shape as we would expect from looking at the large-scale structure. As we reduce the field strength, the cores reflect the overall structure of their parent clouds and become more filamentary. }
    \label{fig:mostmassivecores}
\end{figure*}

\begin{figure}
    \centering
    \includegraphics[width=\columnwidth]{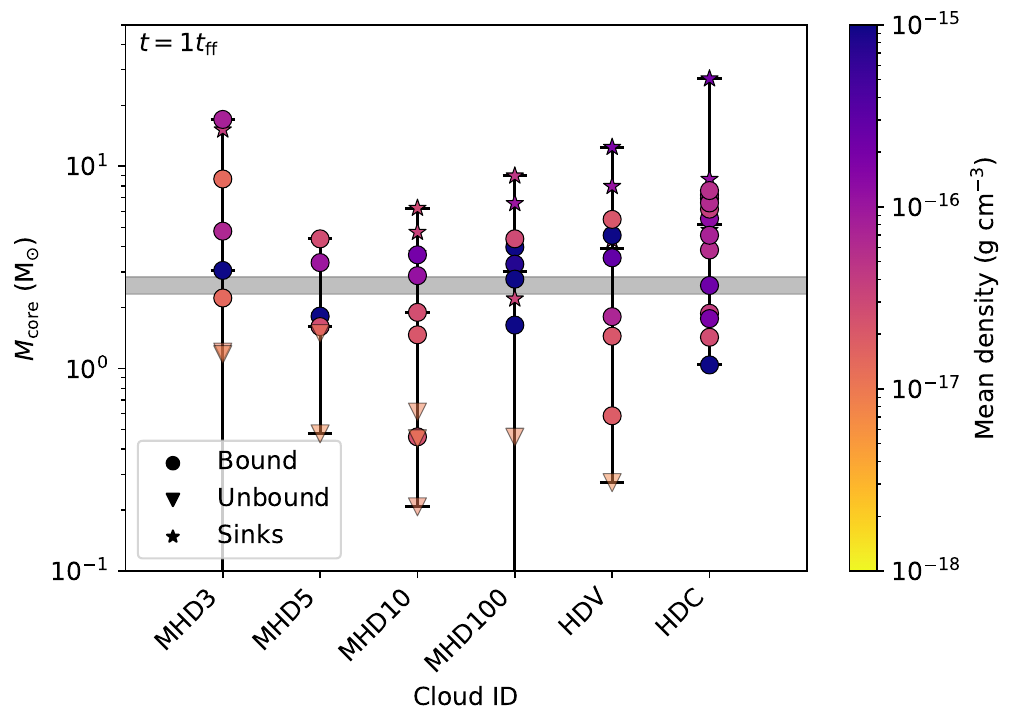}
    \caption{Mass of cores found for each initial cloud magnetization at one cloud free-fall time. The transparent gray area indicates the range of the thermal jeans masses for the clouds, $M_{\mathrm{J}} \approx 2.60\pm{0.25}$ M$_{\odot}$. Transparent, triangular, cores are those which satisfy the energy condition $2T + U \leq 0$, but not $E_{\mathrm{tot}} \leq 0$, and so are not strictly bound. We see these are confined to the very lowest masses, for which $E_{\mathrm{mag}}$ and $E_{\mathrm{thr}}$ are significant. For each set of cores, the maximum, minimum, and median masses are indicated by range bars. Finally, cores with red outlines contain sink particles. Their contribution has also been included in determining the mean core density.}
    \label{fig:core_mass1tff}
\end{figure}

Using the clump-finding algorithm described in section \ref{subsec:CF}, we present the masses of cores identified for each cloud. Figures \ref{fig:corelocscompMHD3HDB} and \ref{fig:mostmassivecores} visualize the cores, whilst figure \ref{fig:core_mass1tff} shows the results for core masses at one free-fall time. In figure \ref{fig:corelocscompMHD3HDB}, we choose to show MHD3 and HDV to demonstrate the two ends of the spectrum for an identical cloud - the strongest field case vs. no field. Figure \ref{fig:mostmassivecores} shows the $x-y$ projection of the most massive gas-only cores in each of the clouds at one free-fall time. We chose not to show cores with sinks in this plot as the influence from the sink can drastically alter the morphology and structure of the surrounding gas. Both of these figures demonstrate how the large-scale differences between the clouds translates to the cores they form - with MHD3 forming more extended and centrally located cores. As the field strength decreases, cores form at increasingly further out locations in the cloud. They are also more noticeably substructured.

From figure \ref{fig:core_mass1tff} we see several trends. With the exception of MHD3, the maximum core mass for each of the clouds increases with decreasing field strength. At this point the difference in maximum mass is less than an order of magnitude; 6.0 M$_{\odot}$ for MHD5 vs 27 M$_{\odot}$ for HDC. We also see that cores switch from unbound (the transparent points) to bound at approximately $\sim 1$ M$_{\odot}$. This is comparable to the initial cloud thermal Jeans Mass, $M_{\mathrm{J}} \approx 2.60\pm{0.25}$ M$_{\odot}$, which approximates the lower limit for fragmentation. This is discussed further in section \ref{subsec:CM_Brho_lm}.

As for the median core masses, we see that they lie broadly around the same value of approximately 1 M$_{\odot}$ for the lowest mass case (MHD5 again), and 6 M$_{\odot}$ for HDC. For the low-mass range, it is once again similar when considering the fully bound cores, i.e., somewhere around 1 M$_{\odot}$, or the original cloud Jeans mass. The total mass within the bound cores is significantly different across the six clouds. For HDC has a total core mass of $M_{\mathrm{c, tot}} = 98$ M$_{\odot}$, followed by MHD3 with $M_{\mathrm{c, tot}} = 50$ M$_{\odot}$. We observe the same pattern as in figure \ref{fig:core_mass1tff}, such that the total mass in cores decreases as the field strength increases (excluding MHD3), such that MHD5 has only 11 M$_{\odot}$ contained in cores by 1$t_{\mathrm{ff}}$. This suggests that core formation in clouds MHD5, MHD10, and to some degree MHD100 is somewhat suppressed with respect to MHD3. These clouds are forming lower-mass cores, and fewer of them, whereas MHD3 forms cores across a range of masses. The reasons for these differences will be further explored in section \ref{subsec:CM_Brho_hm}.

We have three distinct `tiers' of core in the figure, 
Cores shown as stars contain sink particles. In the plot we also include the masses of unbound cores which satisfy $2T + U < 0$, but are not bound when also considering the thermal and magnetic contributions. Whilst they are not currently bound, they may accumulate enough mass at later times and eventually form a bound object. We see that these unbound cores are of the lowest densities and masses.

However, figure \ref{fig:massfracevol} shows that due to the presence of the magnetic fields, the evolution of material within the cloud proceeds at different rates for each of the clouds. Magnetic fields cushion the turbulence, and delay the formation of dense structures which are the seeds for fragmentation. Therefore, comparing the mass of cores at the same physical time (e.g., at one free-fall time) may not provide the best indicator of whether or not a) cores will form in the cloud and b) what properties they have. Figure \ref{fig:massfracevol} shows that for cloud HDC, approximately 1\%, equivalent to 10 M$_{\odot}$, of the cloud mass has passed the core threshold density by a time $\sim 0.8 t_{\mathrm{ff}}$, whereas cloud MHD5 only achieves this at $\sim 0.9 t_{\mathrm{ff}}$. At $t \approx 0.8 t_{\mathrm{ff}}$, MHD5 has less than 1 M$_{\odot}$ available for cores. Therefore, we cannot directly compare the masses of the cores at the same times. For this reason, we use the density evolution of the cloud as a proxy for evolutionary time, and compare the masses and properties of cores when each cloud has the same amount of material above a given density threshold, in this case the threshold core density, $\rho_{\mathrm{thresh}} = 5 \times 10^{-18}$ g cm$^{-3}$.

\begin{figure*}
    \centering
    \includegraphics[width=\linewidth]{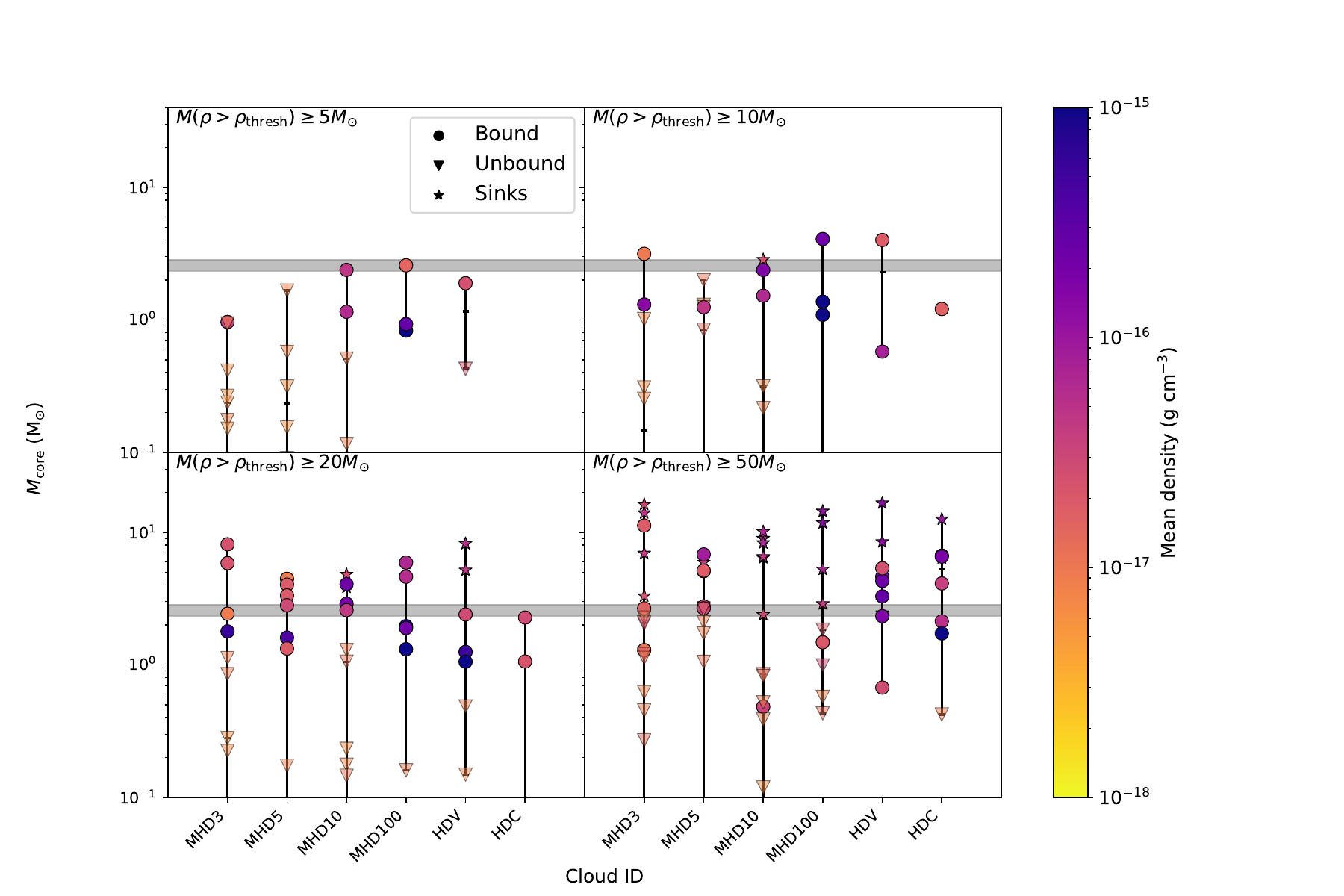}
    \caption{Core mass ranges for comparative points in the cloud density evolution. As before, transparent triangular cores are unbound when considering the contribution from all energy sources, and the thermal Jeans mass range is shown by the transparent grey area. By using density as a proxy for time, we see that the masses of cores for each initial magnetic field no longer have such a large variation in maximum mass, nor in the range of core masses. The mass of cores gradually tends upwards as the snapshots progress, indicating that the cores are not evolving statically, and evolve as a result of their surroundings.}
    \label{fig:core_mass4dense}
\end{figure*}

\begin{figure}
    \centering
    \includegraphics[width=\columnwidth]{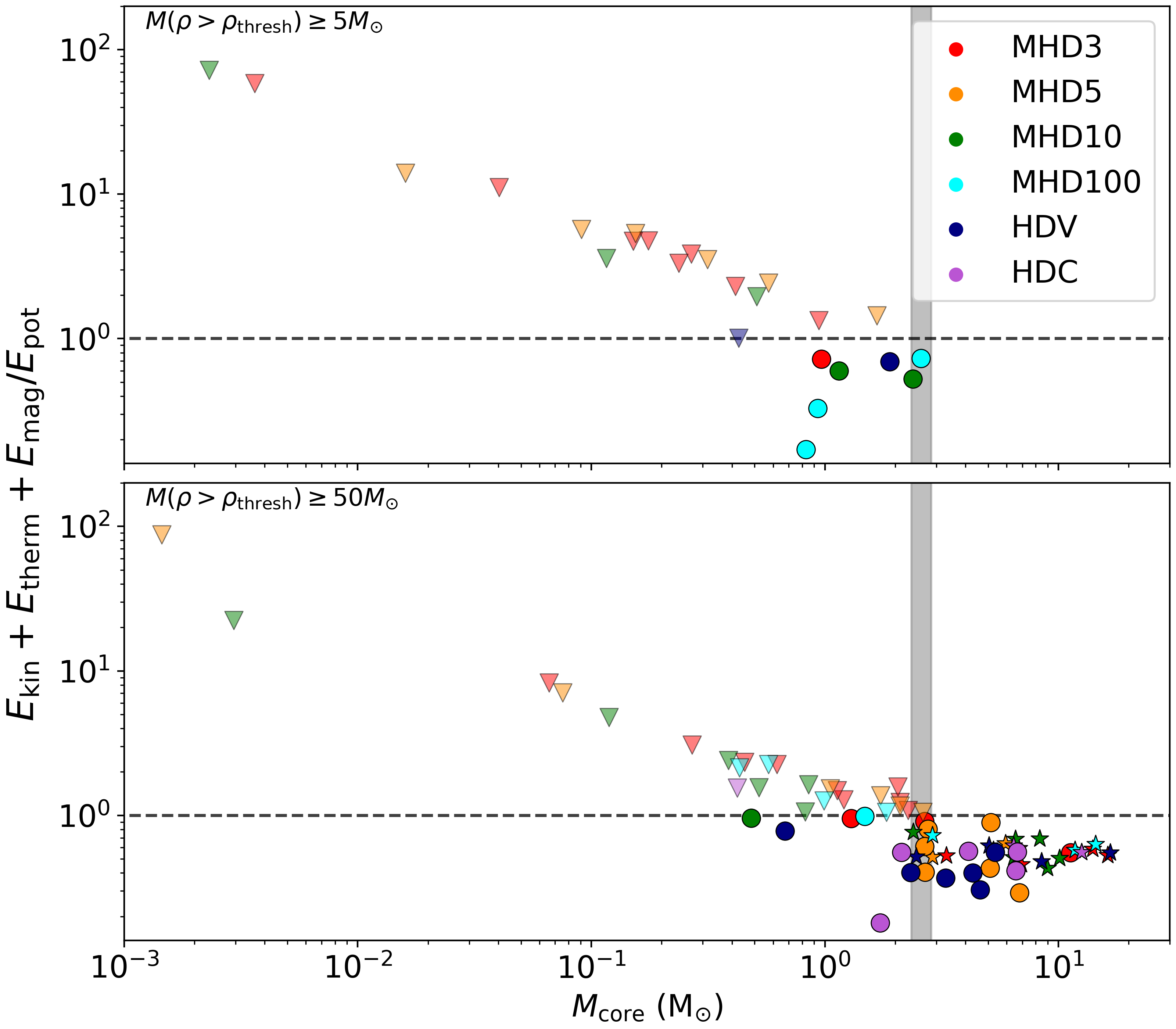}
    \caption{Core energy ratios as a function of their mass for comparative points in the cloud density evolution. Points indicated by transparent triangles are unbound when considering the contribution from magnetic, thermal, and kinetic energies as counters to gravity. We see the energy ratios rise sharply as a function of decreasing core mass, and that in all cases, magnetized or not, cores only become fully bound at masses comparable to the initial cloud Jeans mass (shaded area). As in figure \ref{fig:core_mass1tff}, bound cores are shown by circles, unbound by triangles, and cores containing sinks are shown by stars.}
    \label{fig:MvsEpanel}
\end{figure}

Figure \ref{fig:core_mass4dense} shows the evolution of core masses over four points in density evolution, when $M_{\mathrm{dense}} = $ 5 M$_{\odot}$, 10 M$_{\odot}$, 20 M$_{\odot}$, and 50 M$_{\odot}$ respectively. From this we see several things more clearly. The first is that the masses of bound cores for each cloud are broadly in the same range for each density snapshot, e.g., a $\sim$ 1 - 3 M$_{\odot}$ in panel a), vs $\sim$ 1 - 20 M$_{\odot}$ in panel d). We also see that these ranges shift upwards as the cloud evolution progresses, implying that cores begin at lower masses, and gradually accrete mass as they evolve.

Another feature of the core masses is that the magnetized clouds produce a greater number of unbound, low-mass cores (shown as transparent, triangular, points). As discussed above, whilst these objects may satisfy the energy condition for kinetic and potential energies, they are not necessarily collapsing, due to the additional contribution of thermal and magnetic energies. These objects have the lowest masses and densities. Figure \ref{fig:MvsEpanel} shows the evolution of the ratio of core energies as a function of core mass. From the figure we clearly see that the majority of cores follow a tight correlation from low masses, where their energy ratios are significantly unbound, down to $\sim$ $M_{\mathrm{J}}$ of the initial cloud, where they become bound, and gravity dominates. As all the cores follow this relationship regardless of magnetization, this suggests that magnetic fields are not the most important factor in setting core mass distributions. We will discuss this further in section \ref{subsec:CM_Brho_lm}.

These unbound objects are likely transient, and either disperse as the simulations progress, or, may accrete mass through gas accretion and/or mergers and transition into bound objects at later times. \citet{Weis2024} found a similar result from their simulations of colliding magnetized flows; the suppression of motion perpendicular to the direction of the field causes the cloud material to fragment and form low-, rather than high-mass structures. Similarly to here, the energy ratios of these objects suggest they are transient. This again implies that the formation of higher-mass objects is reliant on the formation of core ``seeds'' which over time accumulate mass from their surroundings, as opposed to the formation of an initially high-mass core. The number of bound cores is broadly similar in all the simulations, and tends to increase as the mass at high density increases. The number of cores ranges from between 1 - 3 in the first few snapshots vs. 2 - 6 at later points. This all suggests core formation which proceeds via the growth of small overdensities into larger structures.

\subsection{Core Mass Function (CMF)}
\label{subsec:CMF}
Another key test for the mode of star formation is to analyse the relationship between the Core Mass-Function (CMF) and the stellar Initial Mass-Function (IMF). If the CMF is populated stochastically, then there should be no relationship between the maximum mass of a core formed in the cluster and the original mass of gas in the clump (e.g., \citet{deWit2005, Parker2007}). Such a result would come from a \textit{Turbulent Core} mechanism of star formation, and would allow for the formation of isolated massive stars. In contrast, \textit{Competitive Accretion} suggests that there should be a correlation between the maximum stellar mass, and the initial mass of gas in the cluster \citep{Bonnell2004, Krumholz2009}. Furthermore, in \textit{Turbulent Core} models of star-formation, we can also expect an almost direct mapping of the CMF into the IMF, as the distribution of mass throughout the cores should remain relatively stable as they proceed through the star-formation process, and we assume an approximately constant star-formation efficiency for each object. Typically, the CMF is approximated to be shifted to 3 times higher masses \citep{Krumholz2009}. In the case of \textit{Competitive Accretion}, the CMF can be expected to evolve much more strongly with time, and at earlier times may be much more bottom-heavy, before cores have had time to accrete mass and grow. \citet{Smith2008} showed that a turbulence core mass-function does not map onto a stellar IMF, due to ongoing fragmentation and accretion.

\begin{figure}
    \centering
    \includegraphics[width=\columnwidth]{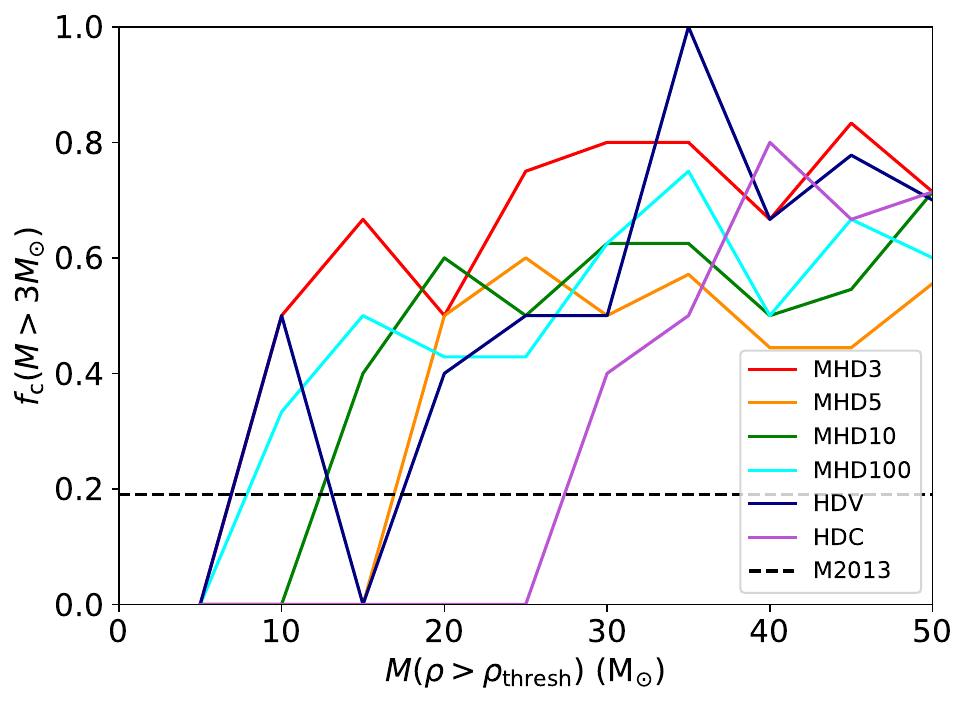}
    \caption{Fractions of bound cores with masses above 3 M$_{\odot}$ as a function of density evolution. Once cores start to form, we see a fairly constant range for $f_{\mathrm{c}}$, between $\sim$ 0.4 - 0.8 in most cases. In all cases this is significantly higher than the value predicted by the CMF approximation using the IMF prescription from \citet{Maschberger2013} (black dashed line).}
    \label{fig:HMFRAC}
\end{figure}

In figure \ref{fig:HMFRAC}, we plot the ratio of bound cores with masses $M > 3$ M$_{\odot}$ as a function of density evolution. We compare this ratio with the fraction predicted when we look at a full IMF. To do this, we take the analytical description of the IMF from \citet{Maschberger2013} (M2013 in the plot legend), and shift the masses up by a factor of three in order to estimate the CMF. We define a break at 3 M$_{\odot}$, and determine the fraction of core masses above this value in the clouds, which would correspond approximately to the break in the IMF at 1 M$_{\odot}$ (e.g., \citet{Miller1979, Chabrier2003}), if we again assume direct mapping. The analytical description originally has a lower limit of 0.1 M$_{\odot}$ (0.3 M$_{\odot}$ for the CMF). However, since no bound cores are found below a mass of $\sim 0.7$ M$_{\odot}$, we truncate the IMF at a value of $\sim 0.23$ M$_{\odot}$, before calculating the fraction of objects above 1 M$_{\odot}$.

For a fully evolved CMF, we get a fraction of $f_{\mathrm{c}}(M > 3\mathrm{M}_{\odot}) > 0.191$, that is, we expect fewer than 10\% of the cores in the sample to have masses greater than 3 M$_{\odot}$. We arrive at this value by sampling the Maschberger IMF 100 times, and taking the mean mass fraction. In contrast, figure \ref{fig:HMFRAC} shows that the clouds show much more top-heavy CMFs. At the earliest times we see a marginally lower value for $f_{\mathrm{c}}$, but as $M(\rho > \rho_{\mathrm{thresh}})$ increases, we start to see $f_{\mathrm{c}}$ stabilize, within a range of $\sim 0.4 - 0.8$ across all clouds.

This suggests that the clouds are broadly producing the same proportion of higher-mass objects regardless of the initial magnetization. However, the strongest field case, MHD3, matches closely with the hydrodynamical cases, whereas the weaker field cases such as MHD5 and MHD10 produce the lowest fraction of objects above 3 M$_{\odot}$. We will revisit this point in section \ref{subsec:CM_Brho_hm}. Despite this, the broad similarities suggest a stochastically filled CMF is unlikely, and that the upper mass limit on the cores is a function of the initial cloud mass. However, as the CMF is not static, it is likely that we must wait to much later times in the cloud collapse for the CMF to begin to resemble the final IMF. 

\section{Discussion}
\label{sec:discussion}

\subsection{Core Masses - Minimum Core Mass}
\label{subsec:CM_Brho_lm}
The initial core masses are likely due to a thermal or magnetic Jeans mass. The thermal Jeans mass is given by
\begin{equation}
    M_{\mathrm{J, thr}} = 1.1 (T_{10})^{3/2} (\rho_{19})^{-1/2} \mathrm{M}_{\odot}
\end{equation}

where $T_{10}$ is the gas temperature in units of 10 K, and $\rho_{19}$ is the gas density in units of $10^{-19}$ g cm$^{-3}$ \citep{Bonnell2006}. For each cloud, we determine the global cloud Jeans mass after one turbulent crossing time, $t_{\mathrm{turb}}$, such that the turbulence has had enough time to impact the internal structures of the cloud. We define $t_{\mathrm{turb}} = R_{\mathrm{Jeans, init}}/v_{\mathrm{turb}}$, where $R_{\mathrm{Jeans, init}}$ is the Jeans length of the initial unperturbed cloud, and $v_{\mathrm{turb}}$ is the velocity of the turbulence, defined by $v_{\mathrm{turb}} = \mathcal{M} c_{\mathrm{s}}$. The only variable which changes $t_{\mathrm{turb}}$ between the clouds is $\mathcal{M}$, so for the four MHD clouds and HDV, $t_{\mathrm{turb}} = 0.18 t_{\mathrm{ff}}$, whereas for HDC, $t_{\mathrm{turb}} = 0.26 t_{\mathrm{ff}}$. We then take the mean cloud density at these times, and calculate the Jeans mass, giving median value of $M_{\mathrm{J, thr}} = 2.60 \pm 0.25$ M$_{\odot}$. This will gradually decrease as the clouds contract.

We can approximate the magnetic Jeans mass following the description from \citet{Li2010,Hennebelle2011}. If $\mu = (M/\Phi)/(M/\Phi)_{\mathrm{crit}}$, then $M_{\mathrm{J, mag}} \approx M_{\mathrm{cloud}}/\mu^{3}$, giving values of $M_{\mathrm{J, mag}} =$ 37.3, 8, 1, and 0.01 M$_{\odot}$ for MHD3, MHD5, MHD10, and MHD100 respectively.

For MHD3 and MHD5, we would expect that the minimum fragment mass is set by the magnetic field, whereas MHD10 and MHD100 should be determined by thermal pressure. We have already seen from section \ref{subsec:masses} that the first cores become bound, and therefore significant, at masses close to the thermal Jeans mass, and this is true for both the magnetized and non-magnetized clouds. 

However, if we consider the mass-to-flux ratio for a fixed volume, such that $R = $ const, then it follows that $\mu \propto \rho/B$, and therefore that $\mu \propto \rho^{1 - \kappa}$. For magnetically dominated collapse, $\kappa$ is small. Material flowing along the field lines acts to decrease the stable magnetic Jeans mass, and cloud fragmentation quickly becomes limited by the thermal Jeans mass. This is further confirmed by figure \ref{fig:MvsEpanel}, in which we clearly see that cores all become bound at around the initial Jeans mass of the cloud, and not at some increased value as a result of the field.

As such, no matter whether the cloud begins in the weak- or strong-field case, the cores will eventually evolve to a magnetically supercritical state in which the thermal Jeans mass is the limiting factor in fragmentation. This broadly explains the similarity in lower-limit core masses across the different clouds.

One caveat from these simulations is that we have not considered the effects of radiative transfer on the evolution of the clouds and cores. \citet{Commercon2010} studied the effect of radiative transfer in combination with magnetic fields for the collapse of a 1 $M_{\odot}$ core, with $(M/\Phi) = 5, 20$, starting with initial temperatures $\sim 10$ K, which is the same as the clouds in this study. Figure 4. from their paper shows that an isothermal approximation for the gas starts to break down at approximately $\rho = 1 \times 10^{-15}$ g cm$^{-3}$, as the gas heats up. This would suppress fragmentation more effectively at higher densities. However, the densities considered within the work presented here are limited to an absolute maximum of $1 \times 10^{-13}$ g cm$^{-3}$, for which figure 4. of \citet{Commercon2010} approximates a temperature of $\sim 30$ K. Furthermore, as can be seen in figures \ref{fig:core_mass1tff} and \ref{fig:core_mass4dense}, the majority of gas-only cores have mean densities in the range $1 \times 10^{-17} - 1 \times 10^{-16}$ g cm$^{-3}$, for which the isothermal approximation still holds. As sink formation begins at densities of $1 \times 10^{-15}$ g cm$^{-3}$, it is unlikely that inclusion of radiative transfer would significantly alter the results of this paper. However, any study of the later stages of core fragmentation would benefit from the inclusion of radiative transfer.

\subsection{Core Masses - Maximum Core Mass Evolution}
\label{subsec:CM_Brho_hm}

The subsequent evolution of the cores differs slightly for magnetized vs. non-magnetized cores. In both figure \ref{fig:core_mass1tff}, and the latter three panels of figure \ref{fig:core_mass4dense}, we see that the most-massive cores in each of the clouds follow an unusual but consistent pattern. The most massive cores are found in MHD3, HDV, and HDC, i.e., the strong magnetic field case, and the two purely hydrodynamical clouds. The lowest masses are found in MHD5, and the rest of the MHD clouds show an increase in the maximum core mass as the initial mass-to-flux ratio decreases. This would suggest that the mechanism by which the cores accumulate mass is different for the most magnetized cases, and the purely hydrodynamical clouds, and that the intermediate magnetization clouds represent a combination of the two modes.

\begin{figure*}
    \centering
    \includegraphics[width=\linewidth]{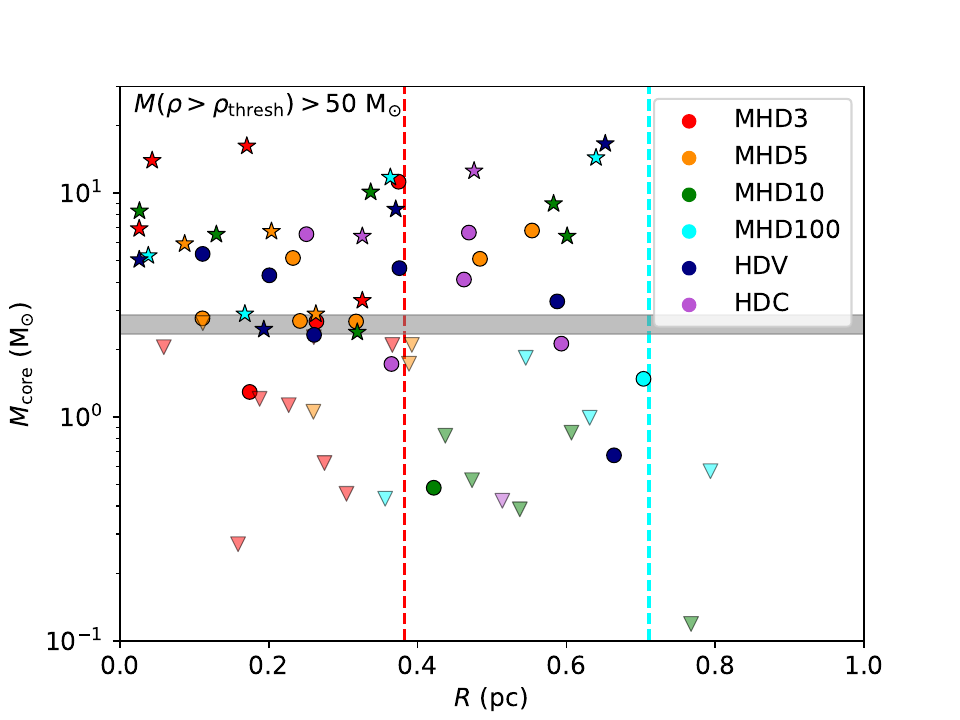}
    \caption{Final core separations from the cloud center are plotted showing that more magnetized clouds produce more centrally located cores. Dashed lines show the maximum extent of bound core locations for the MHD3 and MHD100 clouds respectively. All cores (bound and unbound) from MHD3 are all found within 0.4 pc of the center. In contrast, MHD100, HDV, and HDC all have cores out to a distance of $\sim 0.7$ pc for bound cores, and out to $\sim 0.8$ pc in the unbound case. As before, the Jeans mass range is shown by the shaded grey area, bound and unbound cores are shown by circles and triangles respectively, and cores containing sinks are shown by stars.}
    \label{fig:core_cent}
\end{figure*}

From section \ref{subsec:cloudev}, we see that the distribution of material throughout the clouds differs as magnetization increases. Figure \ref{fig:maxdbins} showed that density peaks formed within the inner 0.5 pc of the most magnetized clouds, and gradually spreads out as the strength of the field decreases. We see this also translates to the locations of cores. Figure \ref{fig:core_cent} shows the distribution of cores for $M(\rho > \rho_{\mathrm{thresh}}) > 50$ M$_{\odot}$. We see a large spread of core locations for the hydrodynamical and weak-field clouds, with MHD100 having the furthest bound and unbound cores from the center, at 0.79 pc, followed by MHD10 at 0.76 pc, and HDV at 0.66 pc. In contrast, the most distant core in MHD3 lies at 0.37 pc.

From this we can begin to understand the two mechanisms by which the cores are forming, and the regimes in which they are important. For strong magnetic fields, the presence of the field acts to erase the turbulent structure, and cloud material flows along field lines and accumulates centrally. Cores form in the central regions of the cloud, and accrete as gas is fed onto the center of the cloud. In comparison, the hydrodynamical clouds form cores much more spread out from the cloud center, due to the overdensities from the initial turbulent seed. Their growth proceeds via some more localized accretion, but cores may also interact and merge with time. Between these two regimes, we have MHD5 and MHD10. The clouds form more substructure due to the turbulence, but on smaller scales the field prevents collapse, meaning the cores cannot grow as large as for MHD100 and the hydrodynamical clouds.

These results are significant as the typical mass-to-flux ratios for star-forming regions are in the range of $(M/\Phi) \sim 3 - 5$ \citep{Crutcher2012, Ching2022}. For the highest field strengths, in line with the lowest mass-to-flux ratios in this study, the growth of high-mass cores is aided by the presence of the field, via mass accumulation along the field lines (e.g., \citet{Ching2022}). Such a process would promote the formation of larger, compressed filamentary structures inside which the cores form. Many studies of star-formation on cloud scales show complex filamentary structures, with filament `hubs' being key sites for massive star formation \citep{Galvan2010, Schneider2012, Hacar2018, Kumar2020}. By comparing the morphologies of cores formed within these simulations directly to observations, we can more concretely describe the processes which lead to the formation of the cores. Furthermore, by directly tracing the evolution of material through back through time, we can understand what processes dominate the growth of cores. These ideas will be explored in more detail in a subsequent paper.

Of course, there is the important caveat that the clouds have been simulated using ideal-MHD conditions such that flux-freezing can be assumed. In reality, the contribution of non-ideal effects would change the relationship of the field to the cloud material. However, as non-ideal effects act to weaken the significance of the field, the overall result that magnetic fields do not produce higher-mass cores than clouds with weak/no magnetic field would be the same.

Furthermore, a full discussion of the CMF is limited by the thermal Jeans mass of the simulations. As we have shown above, the minimum fragment mass is set approximately by this value, independently of the field strength. As such, exploring the full impact of the magnetic field on star-formation statistics would require rerunning the simulations with a much lower value.

\subsection{Magnetic Field -- Density Relation}
\label{subsec:B_rho}

\begin{figure*}
    \centering
    \includegraphics[width=\textwidth]{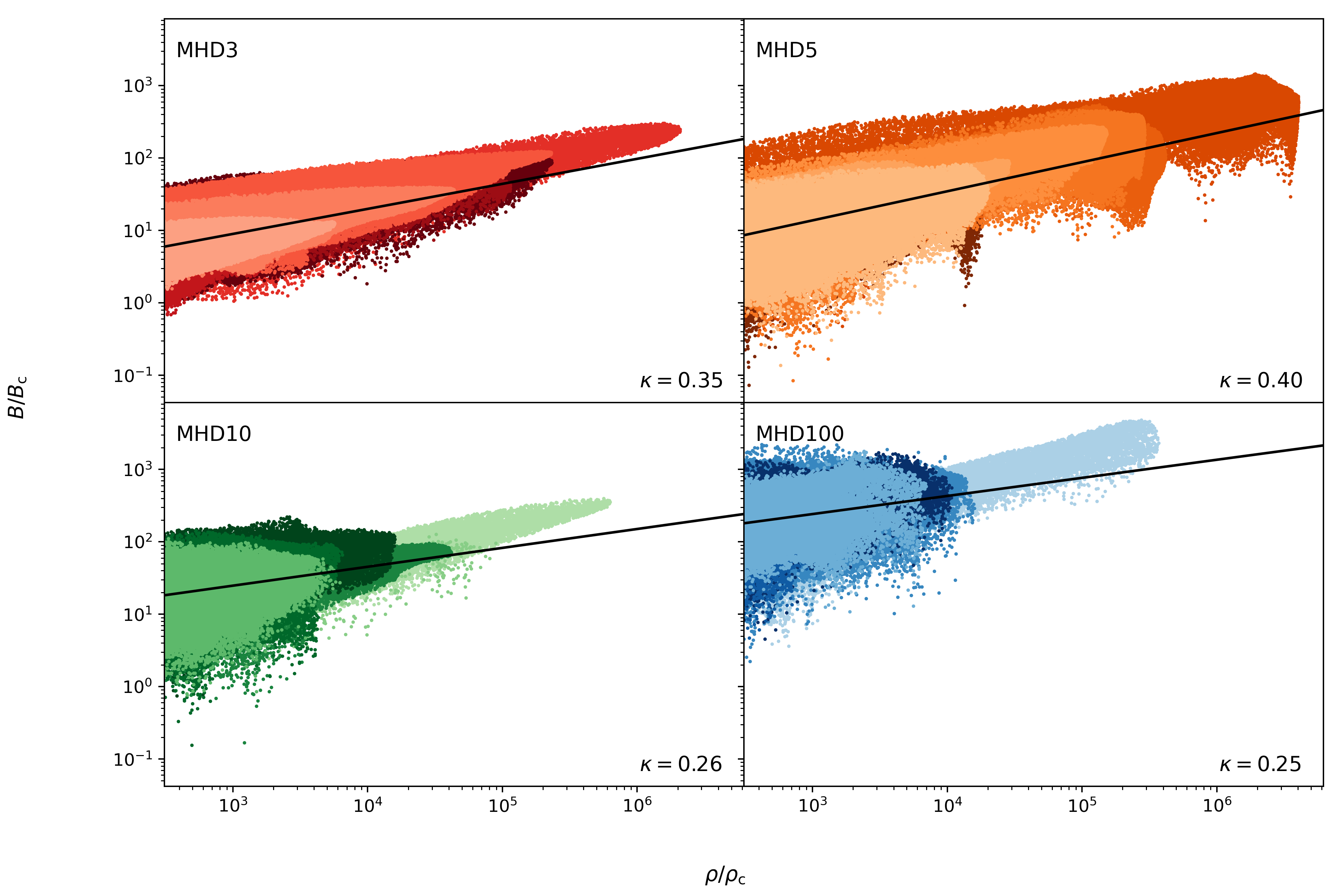}
    \caption{We plot the field strength-density relationship for bound cores within each of the magnetized clouds, at the comparison point $M(\rho > \rho_{\mathrm{thresh}}) > 50$ M$_{\odot}$. Cores are distinguished by the different shades. We have normalized the field strengths by that of the initial B-field of the clouds, $B_{\mathrm{c}}$, and the densities by the initial cloud density, $\rho_{\mathrm{c}}$. Cores from less magnetized clouds have field strengths significantly stronger than the initial cloud magnetization. This is most clear for MHD100, for which the majority of core material has $B \sim 10 - 10^{3} B_{\mathrm{c}}$. We also plot $k$, obtained by fitting each core with a powerlaw, and taking the mean. MHD10 and MHD100 have the smallest values of $\kappa$, implying the field is strongly influencing their collapse. We also see a slight increase in $\kappa$ at the highest densities due to flux freezing.}
    \label{fig:BvRho}
\end{figure*}

To understand why the magnetic field does not appear to affect the masses of the cores which form, we can further analyze the effect of the field on the cloud. Figure \ref{fig:BvRho} shows the evolution of field strength vs density for each of the magnetized clouds. The relationship between magnetic field strength and density can be characterized by a power-law; $B \propto \rho^{\kappa}$, the power of which characterizes the nature of collapse \citep{Mouschovias1976, Trisis2015}. \citet{Mestel1965} first showed that for the collapse of a spherical cloud of constant mass, in flux-freezing conditions, implies a value of $\kappa = 2/3$. However, numerous subsequent observational and numerical investigations of magnetized clumps and cores have returned values ranging typically between $\kappa \sim 1/3 - 2/3$ (e.g., \citet{Mouschovias1976, Crutcher2010, Crutcher2012, Trisis2015, Hennebelle2019}).

There are numerous processes which can affect the value of $\kappa$, but in the case of ideal MHD, where we have flux-freezing, the value of $\kappa$ is set by the morphology of the core that we are looking at. The various possible geometries are discussed in detail by \citet{Trisis2015} in figure 1., but we will summarize the relevant cases for ease. For a spherical core, the mass is related to density and radius via $M \propto \rho R^{3}$, and the magnetic flux through the core is $\Phi \propto BR^{2}$. In the case of flux freezing, the magnetic flux remains constant, as does the mass, and so for a spherical contraction - gravity has overwhelmed the field - we find $B \propto \rho^{2/3}$.
In the case of a disk-like cloud with the field perpendicular to the minor axis (case a in figure 1. of \citet{Trisis2015}), if contraction occurs perpendicular to the field, we find $\kappa = 1$, and the field scales linearly with density. Conversely, if the field is strong enough to counteract gravitational forces (case b), then collapse will only occur along the field lines. The field strength remains constant, as does the area through which the field is passing, and density increases independently of the field, such that $\kappa = 0$. In reality, it is likely to be a combination of contraction along and perpendicular to the field lines, which, as in case c, produces $\kappa = 1/2$.

In reality, the field alignment is likely to be at some angle, $\theta$, to the contraction. \citet{Trisis2015} discuss the evolution of such a cloud with cases f and g, which consider the contraction of cylinders perpendicular to their axis of symmetry. Case f considers small values of $\theta$, such that contraction occurs almost along the field lines, whereas case g considers larger values almost perpendicular to the field. In both cases, they state that as long as there is some contraction perpendicular to the field lines, fragments from the cylinder will form following the relationship $B \propto \rho^{1/2}$.

\begin{figure}
    \centering
    \includegraphics[width=\columnwidth]{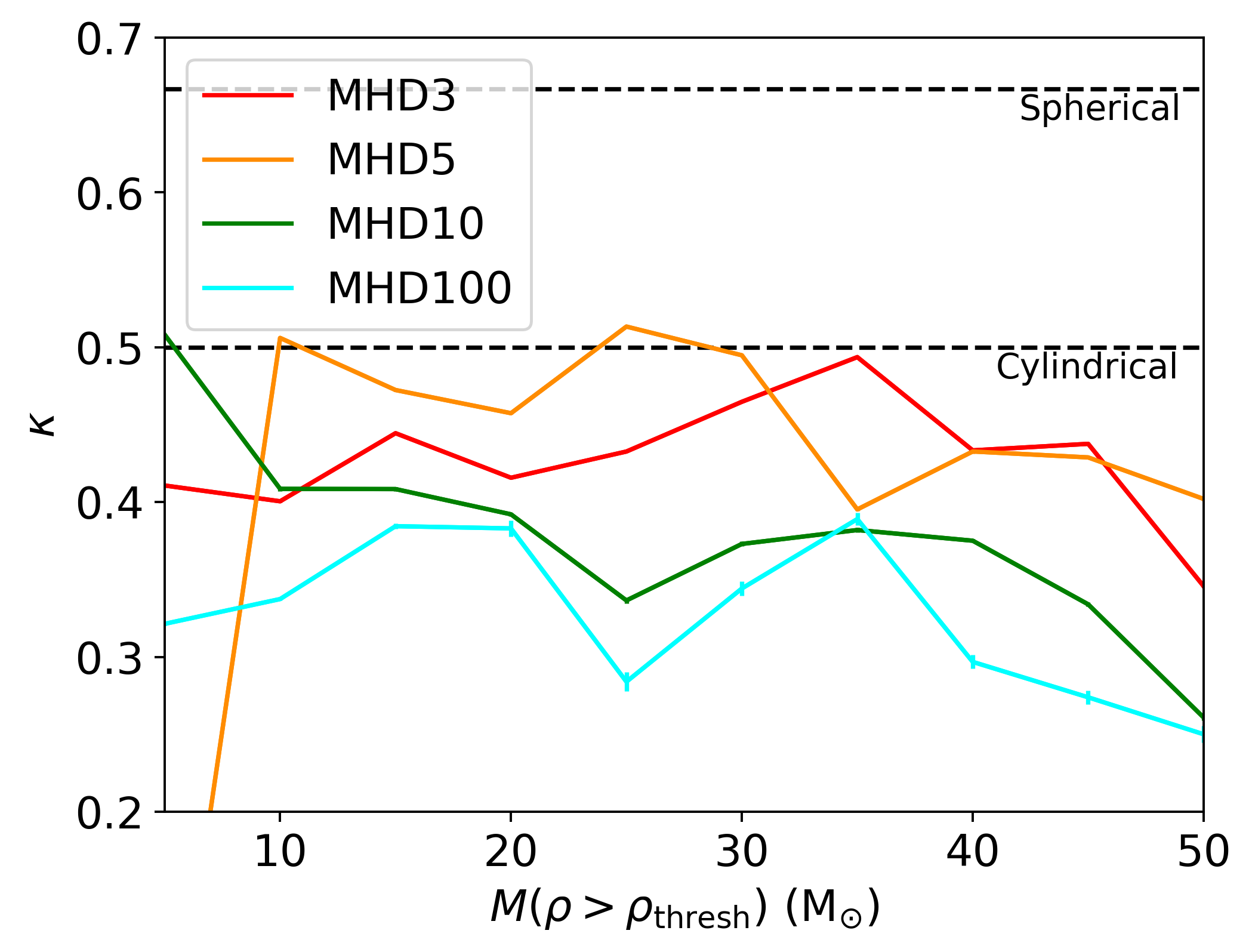}
    \caption{Evolution of mean $\kappa$ in cores as a function of the cloud density evolution. Whilst initially there appears to be no trend, the clouds quickly settle into consistent values of $\kappa$ centered on their respective values at one free-fall time. The clouds appear to show that lowering the initial cloud magnetization also decreases the  $\kappa$. We also see the beginning of a downwards trend at $M(\rho > \rho_{\mathrm{thresh}}) \sim 40$ M$_{\odot}$. Values of $\kappa \rightarrow 0$ indicate material flows along the field lines, suggesting that as the cores evolve the magnetic field becomes mores significant in their evolution. Errorbars are plotted but in most cases are too small to see.} 
    \label{fig:BvRhoevol}
\end{figure}

Figure \ref{fig:BvRho} shows the $B - \rho$ relationship for the cores in each of the four magnetized clouds at one free-fall time. We perform a least-squares fit to the data from each core with a simple power-law function described by 
\begin{equation}
    \log_{10}(B) = \kappa\log_{10}(\rho) + C
\end{equation}
The black line shows the mean value of the power index for each core.

From this figure, we can see two cases emerging. For clouds MHD3 and MHD5, we have a `strong field' case, whereas MHD10 and MHD100 constitute the `weak field cases'. For the strong field cases, the cores follower a steeper power-law relationships, indicating that the field is less dynamically relevant to their collapse than for the weak field cases. We also see that for each initial magnetization, the value of $\kappa < 2/3$, indicating that the cores have more elongated morphologies in each of the clouds.

Figure \ref{fig:BvRhoevol} shows the evolution of this relationship as a function of the density evolution. Very quickly from the point at which the first cores are identified, the values of $\kappa$ settle broadly around the values given in figure \ref{fig:BvRho}. We see that the weaker field clouds have consistently lower values of $\kappa$ indicating the greater importance of the field. The appearance of a downward trend in $\kappa$ at later times suggests that the field becomes more dynamically relevant as the clouds and cores evolve. Whilst this is important for the analysis of core evolution, the focus of this work is to look at the initial masses and properties of cores which form. Furthermore, with the formation of sink particles as the simulations progress, analysis of gas behaviour at later times becomes less robust. As such, we focus on the results from earlier times.

We can understand the decrease in $\kappa$ with weaker initial fields by looking at the field strength values in figure \ref{fig:BvRho}. Plotting $B/B_{\mathrm{c}}$, where $B_{\mathrm{c}}$ is the initial field strength of the cloud, we see that as the initial field strength decreases, the degree of field amplification increases. At densities of $\rho/\rho_{\mathrm{c}} \sim 2 \times 10^{2}$ (corresponding to the core threshold density $\rho_{\mathrm{thresh}}$), material found in cores for MHD3 span $B/B_{\mathrm{c}} \sim 1 - 10^{2}$. As the value of $B_{\mathrm{c}}$ decreases, this range increases, such that for MHD100, the majority of the material in cores have field strengths $\sim 10 - 10^{3}$ times stronger than for the initial cloud. Using the fit values, at the core threshold density, material in cloud MHD3 is predicted to have a field strength of $B = 271 \mu$G. For MHD100, the field strength is $B = 246 \mu$G. Clouds MHD005 and MHD010 have comparable field strengths.

We can understand the relative effect these field strengths have on the cores using the mass-to-flux ratio. As we have seen above, the cores have irregular morphologies, and so calculating the exact area through which the flux is flowing is non-trivial. However, we can provide a rough approximation by stating that $M/\Phi = \rho L/B$, where $\rho$ is the core density, and $L$ is the length of the core along the direction of the field, $B$. We take the $L$ to be the length which encloses 2/3 of the core mass along the mean magnetic field vector in each core, and calculate the density and field strength enclosed by this value. At the final timestep, the cores for each of the magnetized clouds show a range of mass-to-flux ratios spanning as low as $\sim 2$ for at least one core in each sample. At the high end, MHD3, MHD10, and MHD100 have ratios up to $\sim 30$, but MHD5 spans up to $\sim 130$. Although these are rough results, they broadly match the description we see for the field strengths in the cores; regardless of the initial cloud mass-to-flux ratio, the cores evolve to a similar magnetized state.

A full description of the core mass-to-flux ratios, and cloud field-vs-density relationships is beyond the scope of this paper, and will be revisited in the future. However, we can begin to understand this result when considering how cloud material assembles to form cores. In the strong field case; MHD3 and MHD5, the field will slow collapse across the field lines and the cloud will collapse more easily along the field. This leads to a global field-density relationship of $B \propto \rho^{1/2}$. In contrast, lowering the initial field strength allows the global cloud collapse to proceed more uniformally, leading to $B \propto \rho^{3/2}$.

As such, by the time material has evolved to the point of forming cores the support of the field against collapse is the same, regardless of the initial degree of magnetization, explaining the comparable mass-to-flux ratios. This begins to explain why the cores have broadly similar properties across the different levels of magnetization, as well as when compared to purely hydrodynamical simulations.

\section{Conclusions}
\label{sec:conclusions}
We have presented the results from a suite of simulations analysing the effect of magnetic field strength on the formation of high-mass cores. We looked at six 1000 M$_{\odot}$ clouds, each initially supplied with turbulence. Four of the simulated clouds were supplied with magnetic fields such that their mass-to-flux ratios were 3, 5, 10, and 100 respectively. We also analyzed the evolution of two purely hydrodynamical clouds, one of which was sub-virial, such that the cloud collapsed with virtually no support, turbulent or magnetic. The remaining five clouds satisfied the condition $2T + U = 0$. We developed a clump-finding algorithm for use with the 3-dimensional SPH data, and analyzed the clouds at comparable points in their density evolutions. Our main results are as follows:
\begin{enumerate}
    \item The main impact of the field is seen in the global evolution of the clouds. In the strong field cases, the field competes with the initial turbulent seed, and erases the smaller scale structure formation which is seen in the lower/non-magnetized clouds. Material condenses much more centrally, and there is less substructure, especially at larger spatial scales. However, the fields do not prevent collapse and substructures from forming, they just delay the collapse and redistribute the material.
    \item Cores are identified in all of the clouds. Despite the differences in global collapse, all of the clouds show bound cores within similar mass ranges spanning 1 - 20 M$_{\odot}$. We see for all of the clouds that the transition from unbound to bound occurs at $\sim$ 1 M$_{\odot}$, which is comparable to the initial thermal Jeans mass of the clouds. These are significantly different from the magnetic Jeans mass predictions from each cloud, suggesting that the limiting factor in core fragmentation is the thermal Jeans mass.
    \item Regardless of the initial degree of magnetization, the collapse of the cloud amplifies the magnetic field strength such that at core density scales, material is effectively channeled along the field lines, whilst collapse across field lines is more limited. This lowers the magnetic Jeans mass for these densities below the thermal limit, explaining why the comparable masses across the suite of clouds.
\end{enumerate}

\section*{Acknowledgements}
We thank James Wurster for his help with PHANTOM and magnetic fields, as well as his insights for clumpfinding. This project was supported by the STFC training grant ST/W507817/1 (Project reference 2599346). The simulations were performed using the HPC Kennedy, operated by University of St Andrews.

\section*{Data Availability}
Data associated with this paper can be made available upon request to the corresponding author.


\bibliographystyle{mnras}
\bibliography{pap1} % if your bibtex file is called example.bib

%%%%%%%%%%%%%%%%% APPENDICES %%%%%%%%%%%%%%%%%%%%%

\appendix
\label{app}
\section{Core Finding Algorithm}
\subsection{Identifying Candidate Particles}
\label{subsec:candID}
As we are looking for cores in local potential wells, we must first subtract the contribution of the global cloud potential on each gas particle. In order to estimate this value on each particle, we sort them in order of radial distance from the center of the box. We move down each particle in order of distance and, using a spherical approximation for the cloud, the global potential contribution on a given gas particle is given by
\begin{equation}
\begin{aligned}
    E_{\mathrm{pot glob}, i} = -\left(\frac{GM_{\mathrm{enc}, i}m_{i}}{R_{i}} + Gm_{i}\sum_{j = i + 1}^{N - i} \frac{m_{j}}{r_{ij}}\right)
    \label{eq:globpot}
\end{aligned}
\end{equation}
where $E_{\mathrm{pot glob},i}$ is the global potential felt on particle $i$, $M_{\mathrm{enc}} = \sum_{k = 1}^{i - 1} m_{k}$ is the mass enclosed by the particle at a radial distance $R_{i}$ from the center of the cloud, $m_{i}$ and $m_{j}$ represent the masses of particles $i$ and $j$, and $r_{ij}$ is the distance between these two particles. By moving down the particles in order of distance in this way, we avoid performing an $N^{2}$ calculation to determine the global potential on each particle. Whilst this would be a more accurate, it requires significantly more computational time (especially for high $N$) without improving the quality of the results produced by the algorithm. Furthermore, the effect of using a spherical approximation will mostly be felt by particles on the outer edges of the cloud, which are much less likely to be part of cores than those in the center, for which the second term of equation \ref{eq:globpot} will dominate, and therefore yield more accurate global potential estimates.

Once $E_{\mathrm{potglob},i}$ has been determined for all the particles, we subtract it from their total potentials, $E_{\mathrm{pot}, i}$, stored by PHANTOM to determine the local potential,
\begin{equation}
\begin{aligned}
    E_{\mathrm{locpot}, i} = E_{\mathrm{pot}, i} - E_{\mathrm{globpot},i}
    \label{eq:locpot}
\end{aligned}
\end{equation}
Where $E_{\mathrm{locpot}, i}$ is the local particle potential. Finally, we consider for core membership all particles which satisfy
\begin{enumerate}
    \item $E_{\mathrm{locpot}, i} < 0$
    \item $\rho_{i} > \langle \rho \rangle$
\end{enumerate}

\subsection{Initializing Cores}
\label{subsec:initcore}
\begin{figure}
	\includegraphics[width=\columnwidth]{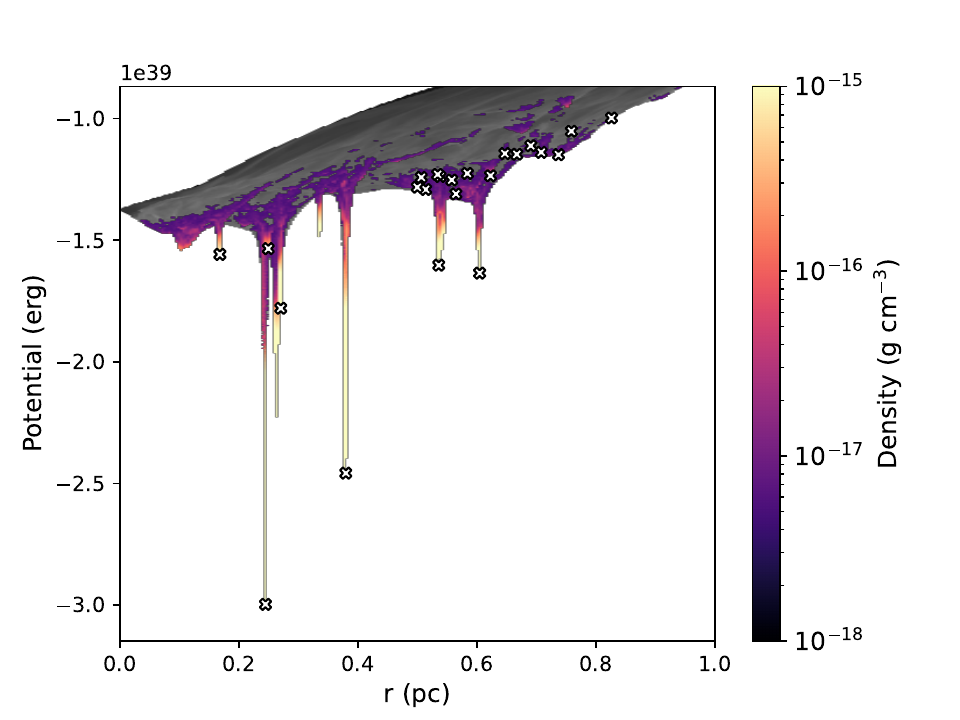}
    \caption{Example of lead sampling in a 1000 $M_{\odot}$ cloud at $\approx$ 0.9 $t_{\mathrm{ff}}$. The full cloud background is plotted in greyscale, whilst the the regions covered by the colour-bar range represent particles which pass the density and potential constraints required for core-membership. The densities represented are the median values at that location. Lead candidate particles indicate the positions of 'seed' particles, from which the core is built up. We see that density peaks do not always correlate with particles in potential wells, and vice-versa, demonstrating the difficulty in comparing cores identified by density vs. potential criteria.}
    \label{fig:rdistpotsamp}
\end{figure}

\begin{figure}
	\includegraphics[width=\columnwidth]{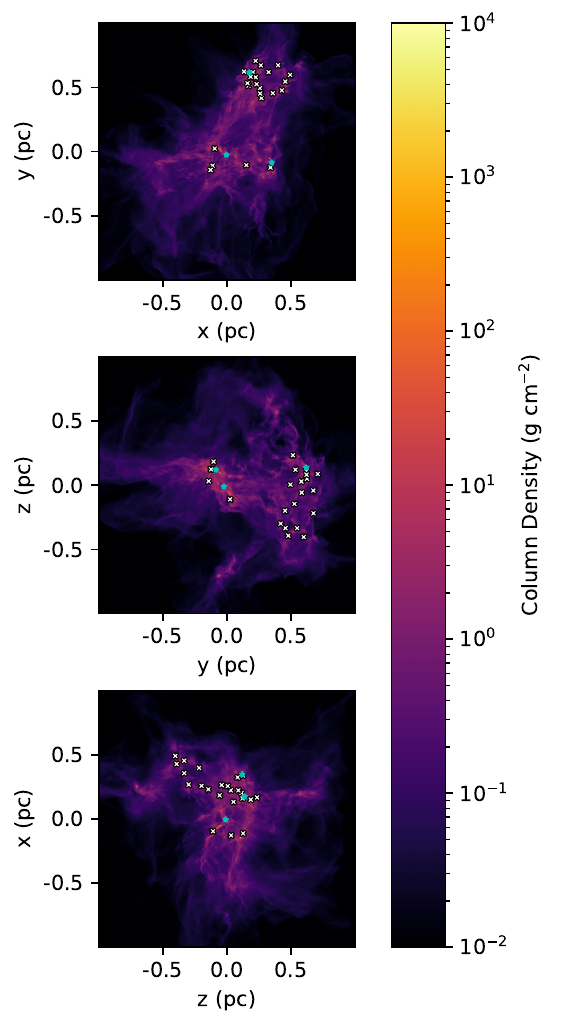}
    \caption{The same cloud as in figure \ref{fig:rdistpotsamp}, shown in all three orientations as a column density plot. As before, the locations of core 'seed' particles are not necessarily well described by peaks in the column density.}
    \label{fig:potsampcloud}
\end{figure}

Next, we want to identify locations of possible cores in the cloud. As discussed above, the simulations contain gas particles, and at later stages of evolution, sink particles are also introduced. Sink particles are automatically assigned as core lead particles within the clump-finding algorithm. For the gas particles, we sort our core candidate particles in order of depth of local potential well. Then, starting from the particle with the deepest well, initialize cores around up to $N_{\mathrm{lead}}$ particles within the cloud. When there are sinks present, we initialize cores around gas particles for $N_{\mathrm{lead}} - N_{\mathrm{sink}}$. To account for possible inaccuracies in the approximation of the background potential subtraction, we also place a core lead particle at the very center of the cloud. In principle this has little effect if the central particle is not in fact the site of a core, as the numerous subsequent checks prevent it from growing erroneously. We chose an upper limit of $N_{\mathrm{lead}} = 100$ starting cores as the number of cores typically converged to a few 10s or fewer.
To avoid placing all of the lead particles within the same local potential well, we also require that all lead particles cannot be too close together. We set this value the lower limit of typical cores sizes of $\sim$ 0.01 pc \citep{Morii2023}, approximately 1\% of the initial cloud radius. Whilst this may seem somewhat arbitrary, we find that the $N_{\mathrm{lead}}$ is usually reached before we reach this minimum separation value. To sample the cloud fully, we start with a large initial minimum separation, and work iteratively down, gradually decreasing the  separation until we have either reached the maximum number of lead particles, $N_{\mathrm{lead}}$, or until the minimum separation distance is reached. We set our initial step size as the minimum value between the initial cloud radius, and the outermost particle considered as part of a core. Figures \ref{fig:rdistpotsamp} and \ref{fig:potsampcloud} demonstrate a typical example of core locations found by the algorithm for a given cloud.


To be assigned as a core lead, gas particles must further meet the following requirements
\begin{enumerate}
    \item The particle density $\rho_{i}$, must be greater than the Roche density, $\rho_{\mathrm{Roche}}$, for the cloud at the particle's location
    \item The particle's value of local potential,  $E_{\mathrm{locpot},i}$ is less than the log-mean value of $ < \langle E_{\mathrm{locpot}} \rangle$ for all our core-candidate particles.
\end{enumerate}
The first condition indicates that the region is self-gravitating against the general cloud environment, and the second condition ensures that we are sampling in the deepest regions of the potential well.

\subsection{Building Cores}
\label{subsec:buildacore}
Now we have identified potential core locations and our candidate particles, we assign particles to cores based whether the particle:
\begin{enumerate}
    \item is within a given distance to the core center-of-mass (COM). This is done so that we avoid running through potential calculations for particles which are unlikely to be part of the core in the first place. We set this distance to the Jeans Radius, $R_{\mathrm{J}} = (M_{\mathrm{J}}/(4/3)\pi \rho)^{1/3}$, of the initial cloud.
    \item is ``connected'' to the core, i.e., the particle smoothing length, $h_{\mathrm{cand}}$, overlaps with the smoothing length of at least one other particle within the core
    \item maintains $2T + U \leq 0$ when added to the core, i.e., the core is bound
    \item maintains $\boldsymbol{\nabla} \cdot \boldsymbol{v} < 0$, i.e., the core is converging
    \item maintains $\rho_{\mathrm{Roche}} < \rho_{\mathrm{core}}$, i.e., the particle is part of the self-gravitating core rather than the overall cloud structure
\end{enumerate}
The particle is checked for candidacy against all active cores simultaneously, and is assigned to the core for which all criteria are met, and to which the particle is most bound.
Once all of the particles have been checked and assigned to cores, we then check whether our cores are overlapping, and if so, whether they should be merged. The merging procedure follows the same requirements as adding particles to the cores.
This process is done repeated until the number of particles in cores has converged.

\newpage
\subsection{Clump-find parameters}
Below is a table of input parameters used for this study, though they can be modified according to the resolution and scale of other simulations.
\begin{center}
\begin{tabular}{ | m{10em} | m{2cm}| m{3cm}|} 
  \hline
  Parameter & Value & Description \\
  \hline
  {\tt nclumpmax}, $N_{\mathrm{lead}}$ & 100 & Sets upper limit on number of cores identified \\ 
  % \hline
  {\tt npartmax } & 2$\times 10^{6}$ & Sets upper limit for arrays which store core particle lists. When this value is surpassed, the algorithm must search through the fill cloud particle list for particle properties, and the program runs slower. \\ 
  % \hline
  {\tt res\_fact } & 116 & Kernel resolution factor \\
  % \hline
  {\tt min\_core\_m } & 116 & Minimum core mass - cores smaller than this value are removed at the end of the clump-finding process \\
  % \hline
  {\tt rhobkg\_cgs } & 1$\times 10^{-20}$ cm & Background density - set to zero if no background \\
  % \hline
  {\tt temp\_gas} & 10 K & Approximate gas temperature\\
  % \hline
  {\tt cloud\_sizecgs } & 3.08 $\times 10^{18}$ cm & Physical extent of cloud - used to define the maximum step size when we are allocate core leads \\
  % \hline
  {\tt minmerg\_sepcgs } & 3.08 $\times 10^{16}$ cm & Maximum separation between two core lead particles considered for merging \\
  % \hline
  {\tt max\_sizecgs } & 3.08 $\times 10^{17}$ cm & Sets maximum size for cores - only necessary when dealing with large particle numbers, as it speeds up the algorithm. \\
  % \hline
  {\tt mincore\_sepcgs } & 3.08 $\times 10^{16}$ cm & Minimum separation required between two lead particles \\
  % \hline
  {\tt rho\_corecgs}, $\rho_{\mathrm{thresh}}$ & 5 $\times 10^{-17}$ g cm$^{-3}$ & Core density threshold \\
  \hline
\end{tabular}
\end{center}

%%%%%%%%%%%%%%%%%%%%%%%%%%%%%%%%%%%%%%%%%%%%%%%%%%

% Don't change these lines
\bsp	% typesetting comment
\label{lastpage}
\end{document}